\documentclass[preprint,aps,tightenlines,floatfix,showpacs]{revtex4}
\usepackage{graphics}
\usepackage{bm}
\usepackage{amsmath}
\usepackage{amssymb}
\begin{document}
\title{Local scale transformations and extended matter distributions
in nuclei}
\author{S. Karataglidis}
\email{kara@physics.unimelb.edu.au}
\affiliation{School  of Physics,  University of  Melbourne,
  Victoria 3010, Australia}
\author{K. Amos}
\email{amos@physics.unimelb.edu.au}
\affiliation{School of Physics, University of Melbourne,
Victoria 3010, Australia}
\author{B. G. Giraud} 
\email{giraud@spht.saclay.cea.fr}
\affiliation{Service de Physique Th\'eorique, DSM, CEA-Saclay, 
F-91191 Gif-sur-Yvette, France.}
\date{\today}
\begin{abstract}
Local scale transformations are made to vary the long range properties
of harmonic oscillator orbitals conventionally used in model structure
calculations of nuclear systems. The transformations ensure that those
oscillator  states asymptotically  have  exponentially decaying  forms
consistent with  chosen single  nucleon binding energies,  leaving the
structure  essentially  unchanged  within  the body  of  the  nucleus.
Application  has been made  to the  radioactive nuclei  $^{6,8}$He and
$^{11}$Li  and  the resulting  wave  functions  are  used to  generate
$g$-folding optical  potentials for  elastic scattering of  those ions
from hydrogen.  As a  consistency test, application  has been  made to
form  wave functions for  $^{40}$Ca and  they have  been used  also to
specify  relevant  proton-$^{40}$Ca   optical  potentials  with  which
elastic  scattering  has  been  predicted. The  need  for  appropriate
specifications of single particle binding energies in exotic nuclei is
discussed.
\end{abstract}
\maketitle


\section{Introduction}

A topic  of current interest is  the description of  the structures of
exotic nuclei, especially as one  approaches the drip lines. The light
mass neutron/proton rich nuclei are particularly suited for study as a
number of these  nuclei can be formed as  radioactive beams with which
experiments to determine their  scattering cross sections can be made.
Their scattering from Hydrogen targets  is of special interest as this
is currently  one of  the best  means by which  the densities  of such
nuclei  may   be  studied  microscopically.  That   is  achievable  as
predictions can be made  of nucleon-nucleus ($NA$) scattering (elastic
and   low  excitation   inelastic)   with  a   folding  model   scheme
\cite{Am00,Ka02},  in  a  manner  consistent with  that  employed  for
electron  scattering. Such allows  for a  sensitive assessment  of the
matter  densities  of nuclei,  as  was  demonstrated  in the  case  of
$^{208}$Pb \cite{Ka02}.  That is the  case also for the  scattering of
radioactive  ions from  Hydrogen,  as inverse  kinematics equates  the
process  to the  scattering  of  energetic protons  from  the ions  as
targets.  However, to make  such predictions~\cite{Am00},  three basic
aspects  of  the  system   under  investigation  are  required.  Where
possible,  these properties  must be  determined independently  of the
proton-nucleus ($pA$) scattering system being studied.

One  must  start with  a  credible  effective (in-medium)  two-nucleon
($NN$) interaction.  Numerous analyses  (to 300~MeV) now  suggest that
such   can  be   deduced  from   $NN$  $g$   matrices;   solutions  of
Bruckner-Bethe-Goldstone  (BBG) equations  based  upon any  realistic
(free) $NN$  potential. With such effective  interactions, analyses of
$NA$ scattering  data become  tests of the  description of  the target
nucleus, namely of its proton and neutron densities.

The second ingredient is the  set of those densities. In the procedure
we adopt,  they are  determined from the  folding of  one-body density
matrix elements (OBDME) and  single particle (SP) wave functions, both
of which  should be obtained  from credible models of  structure. Such
are  normally large  scale structure  models which  describe  well the
ground   state  properties  (and   low-lying  excitation   spectra  if
pertinent) of  the nucleus  in question.

The third ingredient is the specification of the SP wave functions and
it is that  with which this paper is concerned. For  the moment let us
presume that SP wave functions  can be specified appropriately so that
in making a $g$-folding optical potential \cite{Am00} there is nothing
left to be  parameterized as such. When all  elements have been chosen
with care,  namely when appropriate modifications to  the (free space)
interactions between the projectile  nucleon and each and every target
nucleon caused by  the nuclear medium are made, and  when OBDME and SP
wave functions  which well  describe the target  are used  (for stable
nuclei  that  means spectra,  electromagnetic  moments and  transition
rates, and electron scattering  form factors), then predictions of the
scattering  of nucleons  from such  nuclear targets  can be,  and have
been,  made  of  angular  and integral  observables~\cite{Am00}.  That
includes spin-dependent angular  observables. Furthermore, analyses of
data  from  the  scattering  of protons  from  $^{208}$Pb  \cite{Ka02}
clearly  indicated  a preferential  model  of  the  structure of  that
nucleus so  that $^{208}$Pb  should have a  neutron skin  thickness of
0.17~fm.

With  radioactive nuclei,  however, there  are far  less  known static
properties  and no  electron  scattering data  to  complement, and  to
constrain analyses of, the  existent limited hadron scattering data. Of
course structure  models for those nuclei  are a major  field of study
currently and, of note for  the studies we report, several groups have
made shell  model calculations of  the light mass  radioactive nuclei,
$^{6,8}$He    and     $^{11}$Li.    Of    those,     Navr\'atil    and
Barrett~\cite{Na96,Na98}  have made  large-space  calculations (up  to
$6\hbar\omega$  in  the   model  space)  using  interactions  obtained
directly  from  the $NN$  $G$  matrices  which  have the  Reid93  $NN$
interaction   as  their  base.   Also  Karataglidis   \textit{et  al.}
\cite{Ka00} calculated wave functions for $^{6,8}$He within a complete
$(0+2+4)\hbar\omega$ model  space using the $G$  matrix interaction of
Zheng  \textit{et al.}  \cite{Zh95}  based on  the  Nijmegen III  $NN$
interaction.  They   \cite{Ka97}  also  defined   wave  functions  for
$^{11}$Li using  a complete $(0+2)\hbar\omega$ model  space and fitted
potentials.  From  those  wave  functions  the OBDME  to  use  in  the
descriptions   of  both   proton   elastic  scattering   and  of   the
($\gamma,\pi^+$)  reaction (in  the case  of $^6$He  only \cite{Ka00})
were  determined.  Both elastic  proton  scattering  and charged  pion
photoproduction  reactions  probe  the  microscopic structure  of  the
nucleus in a way that initial  states are preserved in the reaction so
that  the  analyses of  scattering  or  reaction  data should  not  be
complicated by the need to describe details of reaction products. With
that assumption, the analyses  \cite{Ka00} confirmed $^{11}$Li to be a
halo nucleus while both $^8$He and $^9$Li are not. The analysis of the
(then) available data on $^6$He did not allow a conclusion on the halo
structure in  $^6$He to  be made. But  the subsequent  measurement and
analysis  of  $p$-$^6$He  scattering  by  Lagoyannis  \textit{et  al.}
\cite{La01}, and later by Stepantsov \textit{et al.} \cite{St02}, have
confirmed that $^6$He has  an extended neutron distribution consistent
with a halo.

Frequently, in  analyses of scattering data,  harmonic oscillator (HO)
wave  functions have  been  chosen to  describe  single nucleon  bound
states in  nuclei. A more realistic representation  may be Woods-Saxon
(WS) functions,  as found for  $^{12}$C \cite{Ka95} for  example. With
the  OBDME   determined  from  $(0+2)\hbar\omega$   shell  model  wave
functions and the single nucleon bound states appropriately specified,
electron scattering  form factors from both the  elastic and inelastic
scattering of electrons from  $^{12}$C then were well fit \cite{Ka95}.
To estimate effects of any halo attribute in the nucleus also requires
variation  of  the  SP wave  functions  from  the  HO set  defined  by
(large-space) shell  model calculations. Such has  been attempted also
using WS  wave functions,  as originally used  in the analysis  of the
strong  $E1$  transition  in  $^{11}$Be by  Millener  \textit{et  al.}
\cite{Mi83}. In  such cases  no constraining electron  scattering data
exist. Even  if there were,  electron scattering data primarily  are a
measure of the proton  distribution of the nucleus. Little information
is obtained directly about the neutron densities from such data.

In  the  case  of a  neutron  halo,  a  specification of  the  optical
potential requires the use of wave functions with the appropriate long
range behavior. Normally,  this is done with the  use of WS functions,
somewhat  artificially. Indeed  to force  a halo  structure  on nuclei
within the traditional (bound state)  shell model, with no coupling to
the continuum,  requires bound state  WS potentials to be  adjusted so
that certain shell model states are weakly bound. A halo structure was
given  to $^{6}$He \cite{Ka00},  for example,  by setting  the neutron
$0p$ shell binding at 2~MeV (near the single neutron separation energy
of  1.8~MeV \cite{Ti00})  and  the  $sd$ shell  and  higher states  at
0.5~MeV  (as dictated  within the  spirit  of the  shell model  single
particle spectrum).  No single  WS potential parametrization  can give
all of those bound states having the relevant binding energies.

However, a  procedure exists that  ensures bound state  wave functions
will   have  asymptotically   an   appropriate  exponential   behavior
\cite{Pe83,St98,Pi01}  whatever  its   originating  form  and  without
sacrificing,  too severely,  bulk  internal character  of shell  model
structure. That involves making  a local scale transformation (LST) of
the  coordinate variable  of the  bound state  wave functions  used in
structure  calculations   (even  if  they  have  been   so  used  only
implicitly).  Namely, given wave  functions which  adequately describe
bulk  properties, we  modify the  tails of  HO SP  wave  functions, as
specified by  the requisite shell  model, for example,  \textit{in the
least artificial way} to  ensure compatibility with whatever choice is
made for single nucleon binding  energies. This is of special interest
for ``halo'' nuclei, or candidates as such.

Herein, Section~\ref{halo} briefly recalls the properties of some such
special nuclei.  Then in Sections~\ref{LST} and  \ref{harm} we explain
the formalism of  the scale transform and give  its justification. The
results of  application of  the LST wave  functions to an  analysis of
proton-nucleus   (nucleus-Hydrogen)   scattering   are  presented   in
Section~\ref{results}. Concluding remarks follow thereafter.

\section{Some aspects of the nuclei $^{6,8}\text{He}$ and
$^{11}\text{Li}$}
\label{halo}

Shell model calculations of $^{6,8}$He and $^{11}$Li have been made to
determine  the   nucleon  shell  occupancies  $n_i$  to   be  used  in
calculations of  the optical potentials for the  elastic scattering of
beams of those ions from  Hydrogen targets. By inverse kinematics that
equates to  proton scattering from  the ions themselves. We  have used
the information from shell model calculations made for earlier studies
\cite{Ka97,Ka00,Am00};  calculations  in  which  all the  nucleons  of
$^{6,8}$He  and $^{11}$Li  were taken  as active  (the  so-called ``no
core''  shell model).  Specifically we  use the  structure information
given  from  those  calculations  of  $^{6,8}$He made  in  a  complete
$(0+2+4)\hbar\omega$  model  space,  and  for $^{11}$Li  made  in  the
smaller  $(0+2)\hbar\omega$ model space.  The latter  space limitation
arose from the dimensionality increasing  with mass for a given space.
While the $^{6,8}$He information came from calculations made using the
$G$  matrix interaction  of Zheng  {\em et  al.}~\cite{Zh95},  the WBP
interaction~\cite{Wa92} was used for $^{11}$Li.

To  utilize  the  LST,  we  list,  in  Table~\ref{spdata},  a  set  of
\textit{estimated}
\begin{table}
\begin{ruledtabular}
\caption{\label{spdata} Estimated binding energies (in MeV) for single
  nucleon orbits in $^{6,8}$He and $^{11}$Li.}
\begin{tabular}{ccccccc}
 Orbit & \multicolumn{2}{c}{$^6$He} &
\multicolumn{2}{c}{$^8$He} &
\multicolumn{2}{c}{$^{11}$Li} \\
 & proton & neutron & proton & neutron & proton & neutron \\
\hline
$0s_{\frac{1}{2}}$ & 24 & 24 & 24 & 24 & 33 & 33 \\
$0p_{\frac{3}{2}}$ & 16.5 & \ 4.0 & 16.5 & 14.5 & 15.7 & \ 7.7 \\
$0p_{\frac{1}{2}}$ & 15.5 & \ 2.0 & 15.5 & 13.5 & 13.8 & \ 5.0 \\
$0d_{\frac{5}{2}}$ & \ 7.0 & \ 2.0 & \ 7.0 & \ 5.0 & \ 2.0 & \ 0.8 \\
$0d_{\frac{3}{2}}$ & \ 5.0 & \ 2.0 & \ 5.0 & \ 4.0 & \ 1.5 & \ 0.8 \\
$1s_{\frac{1}{2}}$ & \ 7.0 & \ 2.0 & \ 7.0 & \ 5.0 & \ 2.8 & \ 0.8  \\
$0f-1p$ & \ 2.0 & \ 2.0 & \ 2.0 & \ 2.0 & \ 0.8 & \ 0.8 \\
\end{tabular}
\end{ruledtabular}
\end{table}
binding energies for nucleons in  the $0s$ to $0f$--$1p$ orbits of the
exotic  nuclei  of interest.  We  stress that  this  set  is used  for
illustration; it should  not be taken as definitive.  In defining this
set we have been guided by the systematics of single particle energies
\cite{Ho94}, on  what WS functions  were needed to match  form factors
from  electron  scattering  from  $^{6,7}$Li  \cite{Ka97a},  and  from
seeking rms  values assessed  from other data  analyses. We  were also
guided  by our  previous  work  involving using  WS  functions in  the
descriptions  of exotic  nuclei \cite{Ka97,Ka00}.  Note also  that the
choice is  dictated by the ordering  of the single  particle states in
the underlying shell  model; this approach differs from  that taken by
Millener \textit{et  al.} \cite{Mi83}, where the  factorization of the
OBDME in terms of spectroscopic  factors connecting to the spectrum of
the $(A-1)$ nucleus make the binding energies change with the relevant
component configurations of the wave function.

In Table~\ref{occups} the orbit occupancies determined from our chosen
shell model calculations,  and up to the $sd$  shell, are listed. With
those occupancies and with a set of SP (proton or neutron) radial wave
functions  $\varphi_i(r)$, we  define  a (proton  or neutron)  density
profile by
\begin{equation}
\rho_{p/n}(r)  = \sum_i  n_i \int  d\Omega  \; \varphi_i^\ast(\bm{r})
\varphi_i(\bm{r})
\end{equation}
where these densities are normalized according to
\begin{equation}
\int_0^\infty \rho_p(r)  r^2 dr = Z  \text{\ \ and\  \ } \int_0^\infty
\rho_n(r) r^2 dr = N \; .
\label{dens-eqs}
\end{equation}
\begin{table}
\begin{ruledtabular}
\caption{\label{occups}
Shell occupancies and rms radii from shell model calculations.}
\begin{tabular}{ccccccc}
 Orbit & \multicolumn{2}{c}{$^6$He} & 
\multicolumn{2}{c}{$^8$He} & 
\multicolumn{2}{c}{$^{11}$Li} \\
 & proton & neutron & proton & neutron & proton & neutron \\
\hline
$0s_{\frac{1}{2}}$  & 1.821 & 1.886 & 1.836 & 1.915 & 1.994 & 1.998 \\
$0p_{\frac{3}{2}}$  & 0.036 & 1.718 & 0.035 & 3.575 & 0.929 & 3.699 \\
$0p_{\frac{1}{2}}$  & 0.036 & 0.262 & 0.038 & 0.329 & 0.037 & 1.474 \\
$0d_{\frac{5}{2}}$  & 0.023 & 0.017 & 0.016 & 0.028 & 0.014 & 0.383 \\
$0d_{\frac{3}{2}}$  & 0.029 & 0.024 & 0.018 & 0.027 & 0.019 & 0.068 \\
$1s_{\frac{1}{2}}$  & 0.031 & 0.034 & 0.035 & 0.036 & 0.006 & 0.373 \\
higher              & 0.024 & 0.059 & 0.022 & 0.090 & 0.001 & 0.005 \\
\hline
$b$ (fm)            & 1.6 & 1.6 & 1.6 & 1.6 & 1.6 & 1.6 \\
$r_{\text{rms}}$ (fm) &  2.11 (2.27) & 2.59 (3.58) & 2.09 (2.20) & 
2.69 (2.79) & 2.16 (2.37) & 2.46 (4.45) \\
\hline
Mass $r_{\text{rms}}$ (fm) & 
\multicolumn{2}{c}{2.44 (3.21)} & 
\multicolumn{2}{c}{2.55 (2.66)} & 
\multicolumn{2}{c}{2.38 (3.99)} \\
\end{tabular}
\end{ruledtabular}
\end{table}
Listed  also are  the oscillator  lengths  used in  those shell  model
calculations and they  lead to the rms radii given  in the second last
line of  the table. The numbers  given in brackets are  the rms values
found using  the LST  functions, that we  define (and  discuss) later,
using the  binding energies in Table~\ref{spdata}. In  the bottom line
we list  the rms  radii for  the entire nuclear  mass, again  with the
values  resulting from  using  the  LST wave  functions  shown in  the
brackets. We consider  first the shell model results  here noting that
the  proton and  neutron  rms  radii differ  for  each nuclei  thereby
naturally identifying a  neutron skin for each. However  the rms radii
obtained  for $^6$He  and $^{11}$Li  do  not define  the neutron  halo
character that  both are  expected to have.  (That will always  be the
case when HO functions are  used.) The proton rms radius obtained from
the LST calculation  is consistent with the oscillator  result in each
case. The neutron rms radii  for $^6$He and $^{11}$Li as obtained from
the  LST  model  are  higher   than  the  oscillator  result  but  are
commensurate  with  those obtained  from  the  WS  and Glauber  models
\cite{Am00}.  There is  agreement in  the neutron  radii  obtained for
$^8$He  from both  the  oscillator  and LST  as  consistent with  this
nucleus being a neutron  skin \cite{Ka00}. The reaction cross sections
for each nucleus are listed in Table~\ref{reacsec},
\begin{table}
\begin{ruledtabular}
\caption{\label{reacsec} Reaction  cross sections (in mb)  at the list
  energy  (MeV) as  obtained  from the  HO  and LST  ($m  = 8$)  model
  calculations.}
\begin{tabular}{cccc}
  Nucleus & Energy & HO & LST \\
  \hline
  $^6$He    & 40 & 321 & 441 \\
  $^8$He    & 71 & 280 & 293 \\
  $^{11}$Li & 62 & 343 & 447 \\
\end{tabular}
\end{ruledtabular}
\end{table}
with the  energies listed reflecting the results  for the differential
cross sections  discussed later.  In the case  of $^6$He, there  is an
experimental  value \cite{Vi02} of  $410 \pm  21$~mb at  36.2~MeV. The
(concocted halo)  WS result at  40~MeV is 406~mb \cite{La01}.  The LST
result of  441~mb remains in  better agreement with these  values than
the HO result  (353~mb \cite{La01}); the slight discrepancy  is due to
the  larger rms radius  compared to  that found  from a  Glauber model
analysis  of  the  interaction   cross  section  [$2.71  \pm  0.04$~fm
\cite{Al96}]. A  similarly small overestimation  in the rms  radius is
observed  for  $^{11}$Li, for  which  the  radius  estimated from  the
interaction cross  section is $3.53  \pm 0.10$~fm \cite{Al96},  and we
expect that  a measurement of  that reaction cross section  would fall
below our  prediction. Nevertheless, we  are encouraged by  the result
for  $^8$He where the  reaction cross  section from  the LST  model is
similar  to the  HO  result as  consistent  with $^8$He  being a  skin
nucleus.

Thus it  is clear that the choice  of SP wave functions  is crucial to
explain observed scattering data and it seems that an important factor
with  that choice  is  the binding  energy  for each  and every  bound
nucleon.  For the  halo  orbits, that  binding  will be  weak and  the
contributions from  those orbits will be small,  commensurate with the
(usually) small occupation numbers  associated with them. Some control
is available by  requiring that the rms radii  be well predicted. Only
with the $^{11}$Li  case is the $1s$ wave  function of some importance
but it  is more significant to have  a form for this  that is extended
noticeably from  the Gaussian  function than it  is to have  a precise
binding energy ---  at least within the context  of the present paper.
As has  been noted \cite{Am00,La01,St02},  it is the reduction  of the
neutron density  within the  core of the  neutron halo nuclei  that is
significant in the analyses of proton scattering. Heavy ion scattering
reflect longer range  properties and so we look forward  to use of the
LST scheme to define density profiles,  etc., that can be used in such
(heavy ion)  reaction studies.  The tabulated values  thus are  a base
input in a study of LSTs to see if the HO functions, used in the shell
model  calculations  to give  the  OBDME,  may  be adapted  to  better
describe  the matter  profiles  and properties  of  these nuclei.  The
present  study of  the LST  is exploratory  and the  calculated matter
densities within this model are not given as ``final'' determinations.

\section{The local scale transformation}
\label{LST}

As given previously \cite{St98,Pi01},  an LST~\cite{Pe83} of the form,
$r =  f(s)$, replaces an  original wave function  $u(r)$ by a  new one
$v(r)$ defined by the isometric transform,
\begin{equation}
v(r) = \sqrt{\frac{df}{dr}}\, u\left[ f(r) \right],
\label{basic}
\end{equation}
where $f(r)$ must  be real and monotonically increasing  when $r$ runs
from 0 to $\infty$. Also, two boundary conditions are in order, namely
$f(0)  =   0$  and  $f(r)   \rightarrow  \infty$  as   $r  \rightarrow
\infty$. The isometry of this  mapping of wave functions $u$ into wave
functions $v$ is obvious  since scalar products are conserved. Indeed,
let  $u$ and $u'$  be two  initial wave  functions and  consider their
respective images $v$ and $v'$ under the transform. Then, trivially,
\begin{equation}
\int_0^{\infty} ds\,  v(s) v'(s) =  \int_0^{\infty} ds\, \frac{df}{ds}
u\left[ f(s) \right] u'\left[ f(s) \right] = \int_0^{\infty} dr\, u(r)
u'(r),
\end{equation}
under  the obvious  change of  integration variable  $r =  f(s)$. With
metrics  for  radial  wave   functions  where  one  uses  an  integral
$\int_0^{\infty}  r^2 dr$, the  transform, Eq.~(\ref{basic}),  must be
slightly modified to,
\begin{equation}
v(r) = \frac{f(r)}{r} \sqrt{\frac{df}{dr}} u\left[ f(r) \right] = s(r)
 u\left[ f(r) \right] \; ,
\label{basrad}
\end{equation}
where $s(r)$ is the wave function scale.

We  are interested specifically  in converting  the usual  shell model
(HO) orbitals  into ones that  have a physical,  exponential decrease.
Let $b$  and $\mu$ denote  the HO length  and the (bare)  nucleon mass
respectively,  and consider  an orbital  that  is bound  by an  energy
$\varepsilon$; the binding  being counted as a positive  number. If we
neglect sub-dominant modulations brought by the polynomials present in
the HO functions and, possibly,  by the derivative $df/dr$, the choice
of $f$ must induce the change in structure
\begin{equation}
\exp{\left(   -\frac{r^2}{2b^2}   \right)}   \Rightarrow   \exp{\left(
-r\sqrt{\frac{2\mu\varepsilon}{\hbar^2}} \right) } \; .
\end{equation}
Hence, when $r \rightarrow \infty,$ we must constrain $f$ by 
\begin{equation}
f(r)  \rightarrow \gamma  \sqrt{r}, \text{  with }  \gamma =  b \left[
\frac{8 \mu \varepsilon}{\hbar^2} \right]^{\frac{1}{4}}.
\label{gamdef}
\end{equation}
Simultaneously,  it seems  best to  set $f(r)  \rightarrow r$  when $r
\rightarrow  0$.  This choice  leaves  the  interior  of the  orbitals
essentially  unchanged.   Accordingly,  the  transition   between  the
``inner,  intact''   regime,  $f(r)  =  r$,  and   the  ``outer,  tail
compatibility'' regime, $f(r) = \gamma \sqrt{r}$, must occur about the
point $r = r_t = \gamma^2$,  which we define as the transition radius.
But a  choice will  need to be  made between two  solution conditions,
namely
\begin{description}
\item[(i)] extension of the linear  regime to respect the initial wave
function as much as possible, and
\item[(ii)]  fix  the  transition   according  to  the  SP  separation
energies,  as  soon  as  $r$  is  of the  order  $\gamma^2$  for  each
individual orbital.
\end{description}

Geometrically, condition  (i) consists in keeping a  straight line for
$f(r)$, overshooting the $\gamma  \sqrt{r}$ parabola, then bending the
formerly straight line slowly to reach the parabola from upper values.
The  second  choice  consists  of an  unbiased  interpolation  between
straight line  and parabola, thus deviating earlier  from the straight
line. In that case $f(r)$ will  always lie below both the line and the
parabola limits  and its derivative will remain  positive definite and
monotonically  decreasing.   Thus  under  under   condition  (ii)  the
normalization in  Eq.~(\ref{basrad}) is always real  and the transform
gives a new function that gains the larger orbit probability amplitude
at  long  range (exponential  rather  than  Gaussian)  at the  expense
primarily of  the surface region Gaussian amplitudes.  We believe that
the  condition  (ii) features  are  sensible  ones  to have  with  the
transform,   especially   as   a   negative  gradient   (and   thereby
indeterminate normalization) is not prevented with condition (i).

\section{The harmonic mean form}
\label{harm}

Solution condition (ii) is met if we use a harmonic mean form for  the 
LST, namely
\begin{equation}
f(r)  =  \left[  \frac{1}{   \left(  \frac{1}{r}  \right)^m  +  \left(
\frac{1}{\gamma\sqrt{r}} \right)^m } \right]^{\frac{1}{m}}\ .
\end{equation}
This  form has  the  added  attractive character  in  that it  depends
primarily upon the chosen SP  binding energies. The order $m$ controls
how sharply the transform alters the coordinate between the limits. We
present,   empirically,  the  results   for  $f(r)$,   its  derivative
$\frac{df(r)}{dr}$ and of the  wave function scale $s(r)$ for harmonic
mean forms  with $m =  4$ and  8 in the  next two figures.  These test
calculations were  made using $b  = 1.6$~fm and  a mass of 1.  In both
figures, the results shown by the solid and dashed curves respectively
are for binding energies of 1  and 20~MeV. In the top segment of these
figures,  the  dot-dash  curves  display  the parabolas  $f  =  \gamma
\sqrt{r}$ for each orbital while  the dotted line is the central limit
of $f =  r$. In the middle segment of each  figure, the derivatives of
the  transformations are displayed.  In the  bottom panel  the scaling
function  with  which  the  transformed wave  function  $u\left[  f(r)
\right]$ is multiplied in Eq.~(\ref{basrad}) is shown.

\begin{figure}
\scalebox{0.6}{\includegraphics*{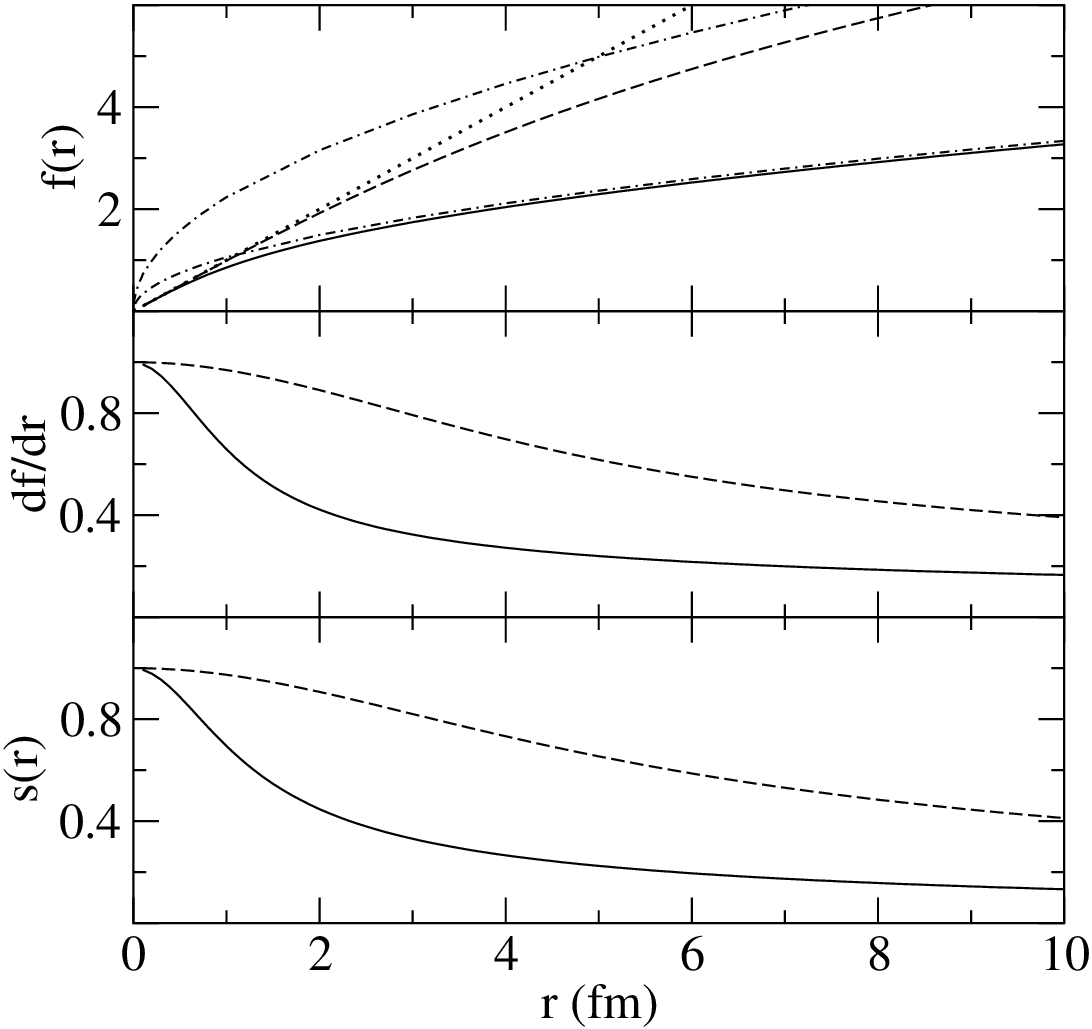}}
\caption{\label{example-m4}  The  $m  =  4$  harmonic  mean  transform
results for  binding energies  of 1 MeV  (solid curves)  compared with
those  for a  binding  energy of  20  MeV (dashed  curves). The  limit
transform  functions also are  shown in  the top  panel by  the dotted
curve [$f(r)  = r$] and by  the two dot-dashed curves  [$f(r) = \gamma
\sqrt{r}$].}
\end{figure}

\begin{figure}
\scalebox{0.6}{\includegraphics*{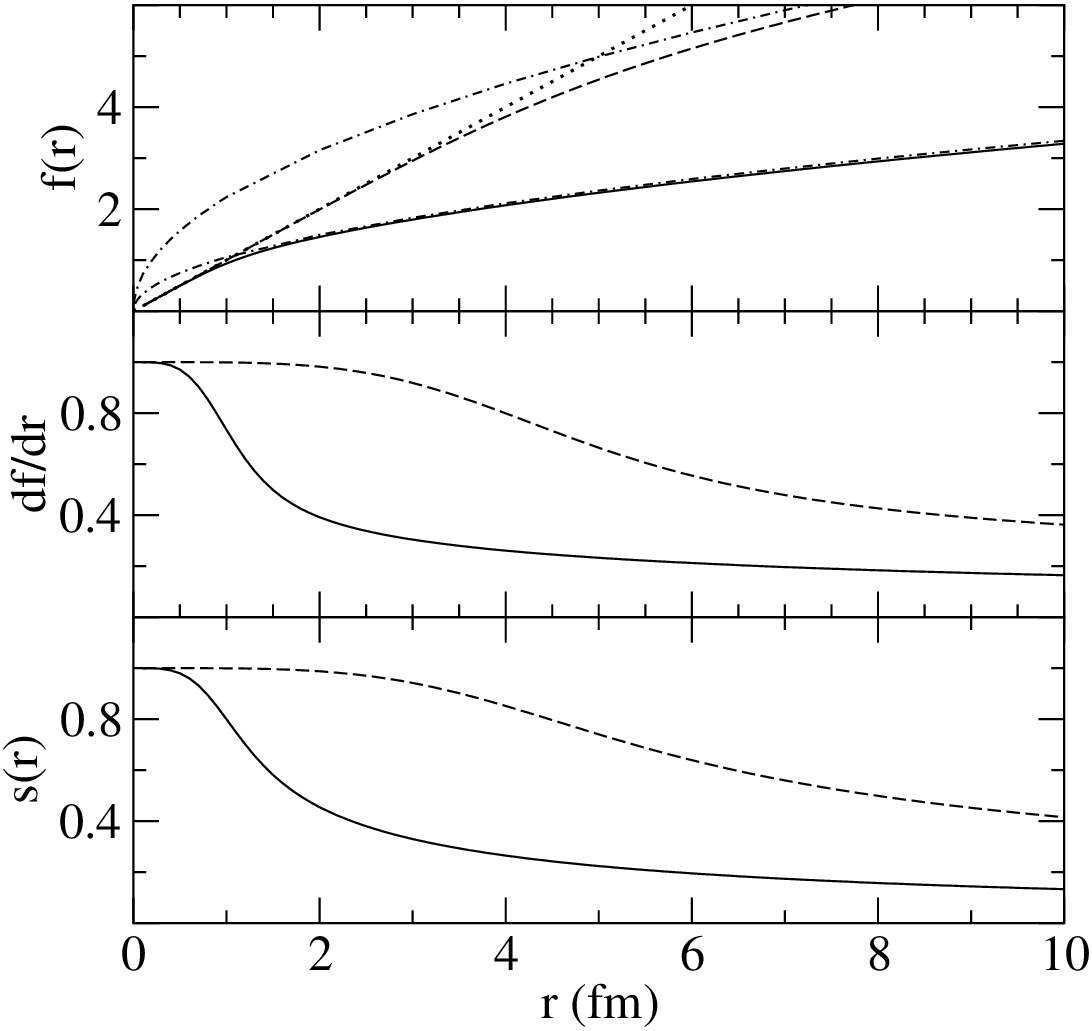}}
\caption{\label{example-m8}  The  transform   functions  as  given  in
Fig.~\ref{example-m4}, but now for $m = 8$.}
\end{figure}

For both  the $m = 4$  and $m =  8$ harmonic mean cases,  portrayed in
Figs.~\ref{example-m4}  and  \ref{example-m8}  respectively, a  weaker
binding induces an earlier transition from the linear to the parabolic
regime.  The  derivatives  also  vary monotonically  to  give  scaling
functions  that do  so  as well.  (For  the sake  of completeness,  we
investigated several values of $m$, of which  $m = 4$ and $m = 8$ only
were chosen  for the  figures.) Therein it  is readily seen  that with
larger $m$ the interpolating curve follows initially the straight line
limit from  the origin before  smoothly, but more quickly,  varying as
the parabola. Actually, if $m \rightarrow \infty$, then $f(r)$ becomes
strictly linear  until the intersection point between  the two regimes
and  then  strictly  parabolic  beyond.  At this  limit  however,  the
continuity in the derivative of $f$  is lost. That loss would make our
transformed  wave function  have a  discontinuity as  well, and  was a
reason for  our choice of moderate  values of $m$  for calculations of
the  nuclear wave  functions to  be used  later.

The  key role  played by  the binding  energy in  modulating  the wave
functions  is  apparent  from  these  diagrams as  well.  Besides  the
transform  effect   of  changing  Gaussian   radial  distributions  to
exponentials  with the  appropriately defined  exponents,  the scaling
functions  depicted in  the  bottom segments  show  that, with  deeper
binding, the interior  character of a shell model  wave function would
be retained more  than those for weaker binding.  Also the increase of
power (from  $m =  4$ to 8)  causes the  variation to be  more surface
oriented. That is  a consequence of the transform  remaining closer to
the linear limit until the break point, which increases in radius with
binding energy  (larger $\gamma$).  It is important  to note  that the
normalising  scale  function, $s(r)$,  tends  slowly  to  zero as  $r$
increases which has a consequence for the densities obtained.

We  show  now  the  cases  for three  exotic  nuclei,  $^{6,8}$He  and
$^{11}$Li.  The last,  $^{11}$Li,  is a  special  case as  we need  to
address  with it  a question  of non-orthogonality.  It, of  the three
exotic light mass nuclei considered, has a sizable $1s_{\frac{1}{2}}$
neutron shell occupancy, as consistent with its $s$-wave halo.

\subsection{The case of $^6$He}

\begin{figure}
\scalebox{0.6}{\includegraphics*{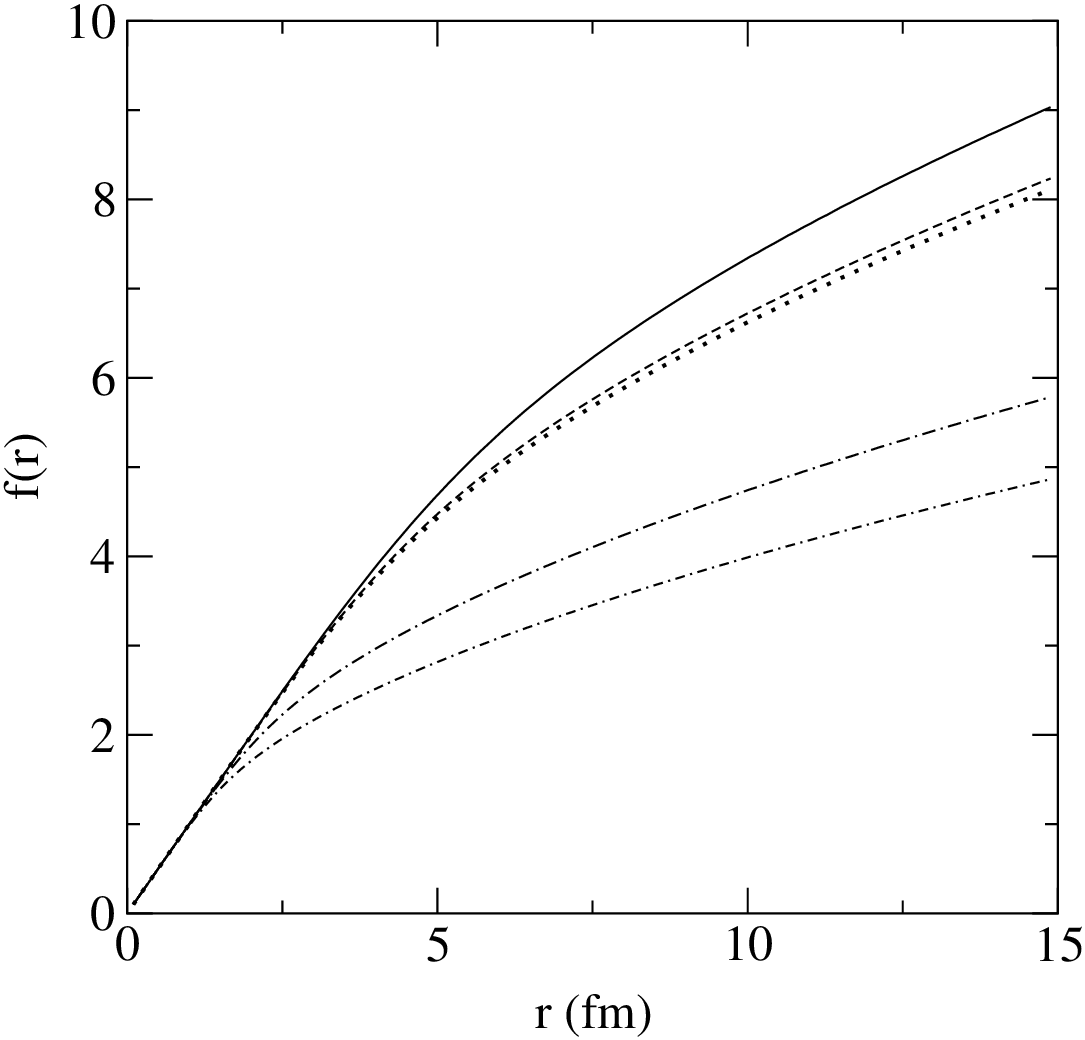}}
\caption{\label{He6-8hm-ftran}  The  $m=8$  harmonic  mean  coordinate
transform functions for  the orbits of $^6$He found  using the binding
energies  listed   in  Table~\ref{spdata}.  The   transforms  for  the
$0s_{\frac{1}{2}}$, $0p_{\frac{3}{2}}$, and $0p_{\frac{1}{2}}$ protons
are shown from the top by  the solid, dashed, and dotted curves, while
those of $0p$-shell neutrons bound by  4 and 2~MeV are depicted by the
dot-dashed and double-dot-dashed lines respectively.}
\end{figure}

\begin{figure}
\scalebox{0.6}{\includegraphics*{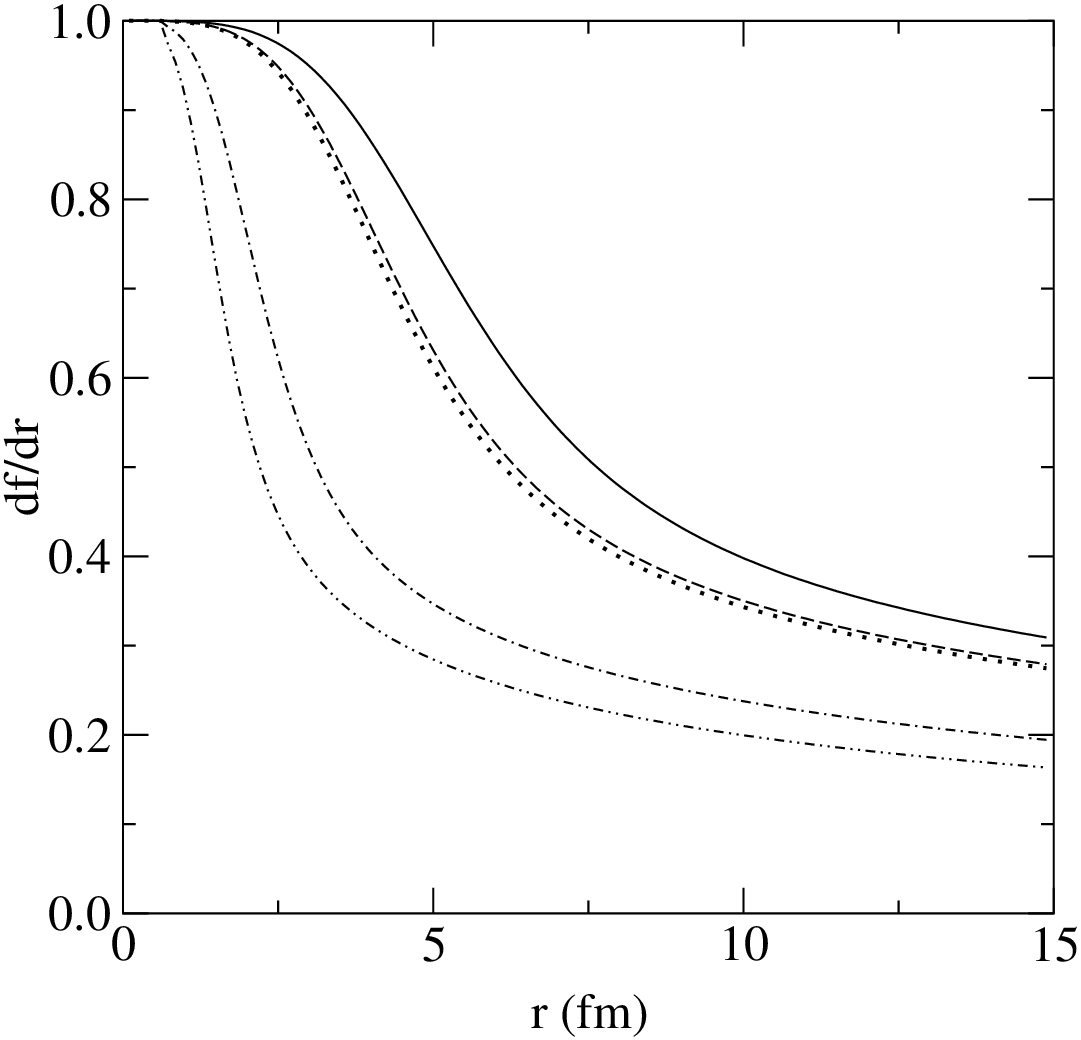}}
\caption{\label{He6-8hm-diff} As  for Fig.~\ref{He6-8hm-ftran} but for
the derivatives, $\frac{df(r)}{dr}$.}
\end{figure}

\begin{figure}
\scalebox{0.6}{\includegraphics*{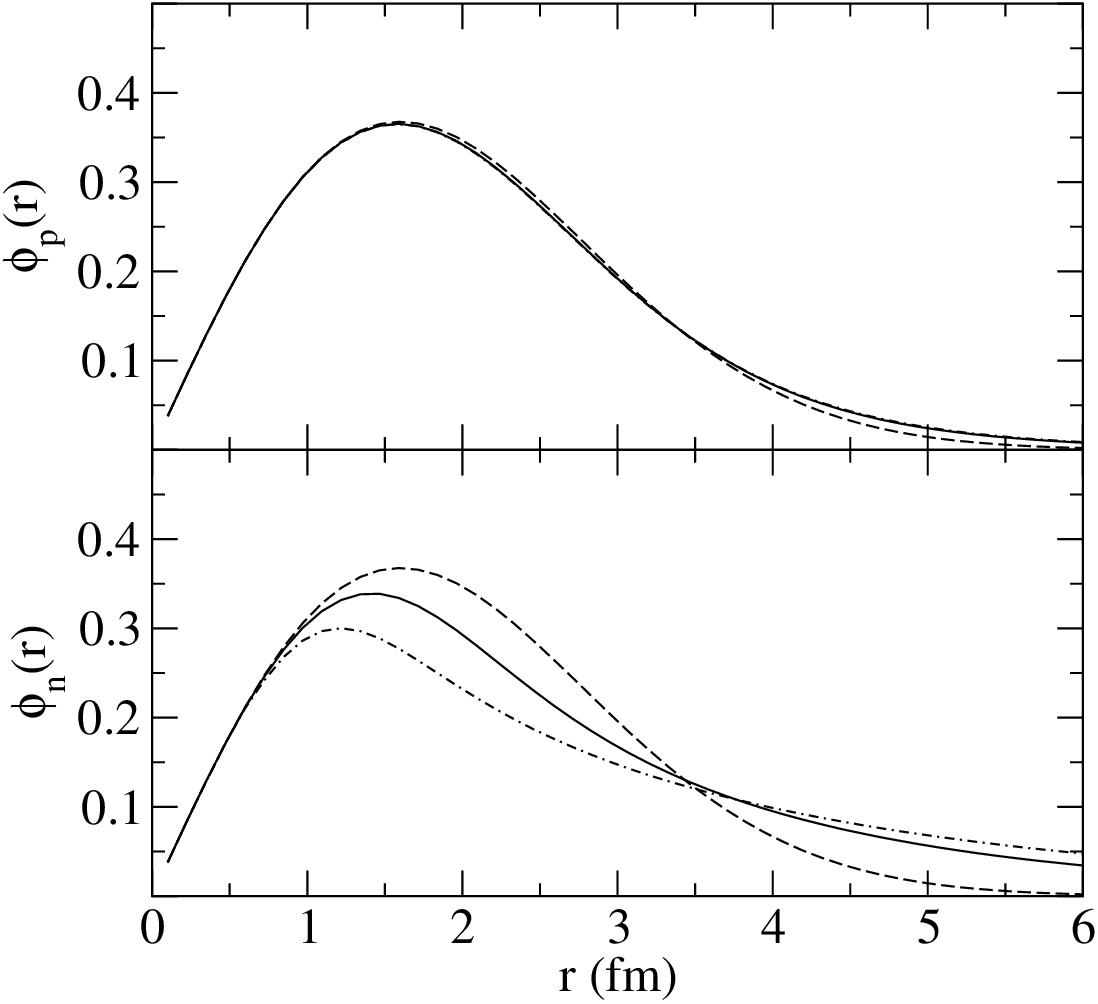}}
\caption{\label{He6-hmean-wvfns}  The  $p$-wave  orbit  functions  for
$^6$He. The  LST results are  shown by the  solid ($0p_{\frac{3}{2}}$)
and dot-dashed ($0p_{\frac{1}{2}}$)  curves for various binding energy
values as stated in the text  while the HO wave functions are shown by
the dashed curves.}
\end{figure}

\begin{figure}
\scalebox{0.6}{\includegraphics*{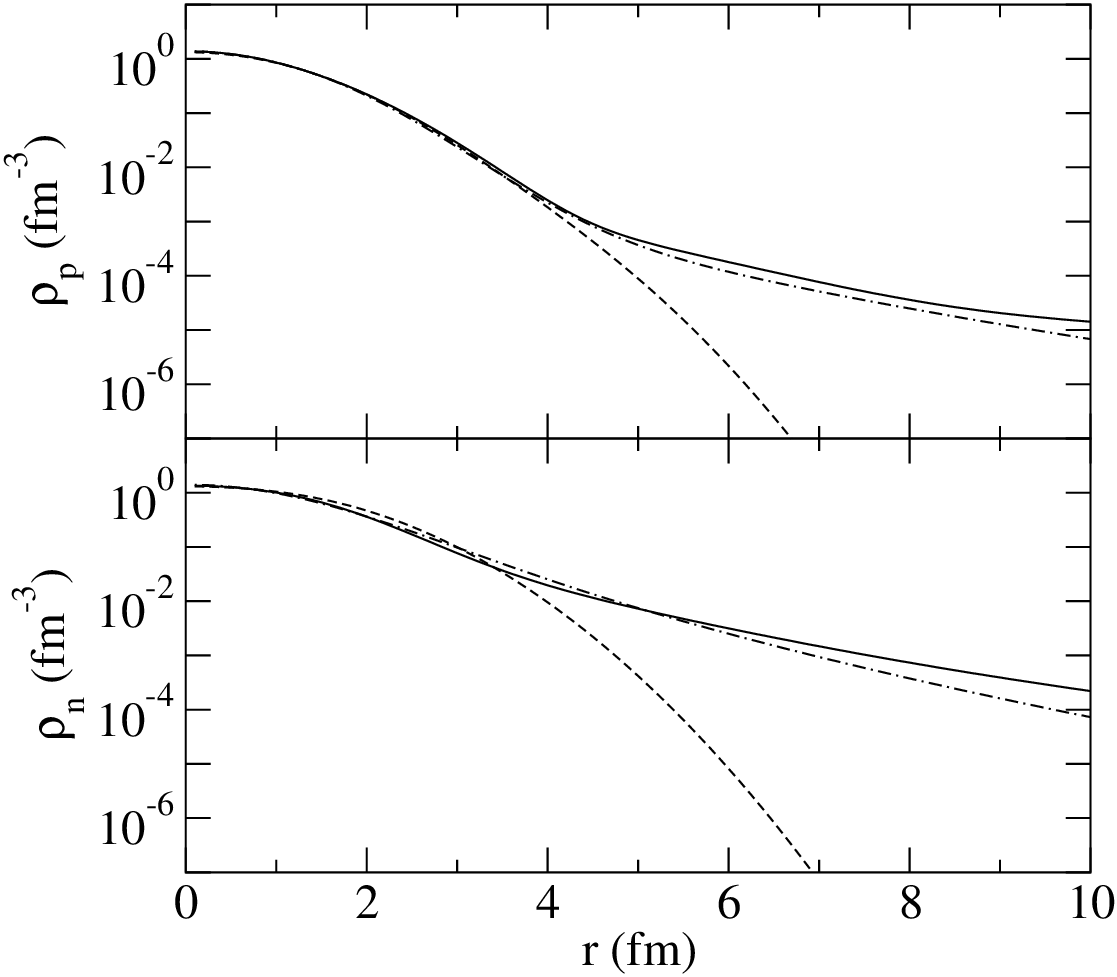}}
\caption{\label{He6-8hm-dens}  The proton  (top) and  neutron (bottom)
densities for $^6$He. The HO, LST, and WS results are portrayed by the
dashed, solid and dot-dashed curves, respectively.}
\end{figure}

In Figs.~\ref{He6-8hm-ftran}, \ref{He6-8hm-diff},
\ref{He6-hmean-wvfns}, and \ref{He6-8hm-dens},  we show the
harmonic mean ($m  = 8$) results for the  transformation functions and
their derivatives,  the individual  wave functions, and  the densities
for $^6$He.  In Figs.~\ref{He6-8hm-ftran} and  \ref{He6-8hm-diff}, the
coordinate   transform   functions   $f(r)$  and   their   derivatives
$\frac{df(r)}{dr}$ for the $0s_{\frac{1}{2}}$ through $0p_\frac{1}{2}$
wave  functions  for  $^6$He  with  the  binding  energies  listed  in
Table~\ref{spdata},  are shown.  The identification  of  the different
orbit results is  given in the figure caption.  Since the scale factor
is  quite  similar to  the  derivative  functions  for most  radii  of
interest  they  are  not  shown.  However,  from  the  shapes  of  the
derivatives, the  $0s$ orbit will remain  essentially unchanged inside
the nuclear  volume, such as it  is for $^6$He, while  the $0p$ orbits
will be influenced more, especially  the neutron orbits. The degree to
which this is the case is shown in Fig.~\ref{He6-hmean-wvfns}. The top
panel  gives two $0p$  wave functions  generated from  that oscillator
function using the $m = 8$ harmonic mean LSTs with binding energies of
of Table~\ref{spdata}. Note that for  the protons (top panel) there is
only  a very  slight change  to effect  the exponential  forms  with a
reduction of  the amplitudes  for radii  in the range  1 to  3~fm. The
weaker bound  (neutron) orbits  in contrast are  much varied  from the
starting HO form with a reduction through the nuclear interior to give
the strong enhancement asymptotically. Thus an extended neutron (halo)
distribution can  be formed by  summation over the  orbit occupancies.
The  results  are shown  in  Fig.~\ref{He6-8hm-dens}  with proton  and
neutron matter densities in the top and bottom segments, respectively.
The HO,  WS, and LST results  are portrayed by  the dashed, dot-dashed
and  solid   curves,  respectively.   The  neutron  halo   is  clearly
established by both  the WS and LST model results  as compared to that
from the  HO model.  Note that the  asymptotic properties of  the wave
functions, and therefore densities, tend slowly towards an exponential
form.  From the  LST,  this is  due  to the  behavior  of the  scaling
function for each orbital, $s(r)$, tending to zero only as $r^{-3/4}$.
The  consequence  of that  extension  in  the  neutron density  is  an
extension also  for the  proton density, though  not quite  as strong.
This stems  from the  addition of small  contributions from  the loose
binding  in the  proton SP  orbits  folded with  the small  occupation
numbers for the  proton orbits above the $0p$ shell,  all of which are
of  comparable  size.  That  dilution  of the  proton  density  by  an
extensive neutron  density, due to the  effects of the  $NN$ force, is
expected in  heavy, neutron-rich, nuclei. This is  consistent with the
slightly  larger proton  rms radius  obtained from  the LST  model, as
compared to the oscillator result.

\subsection{The case of $^8$He}

\begin{figure}
\scalebox{0.6}{\includegraphics*{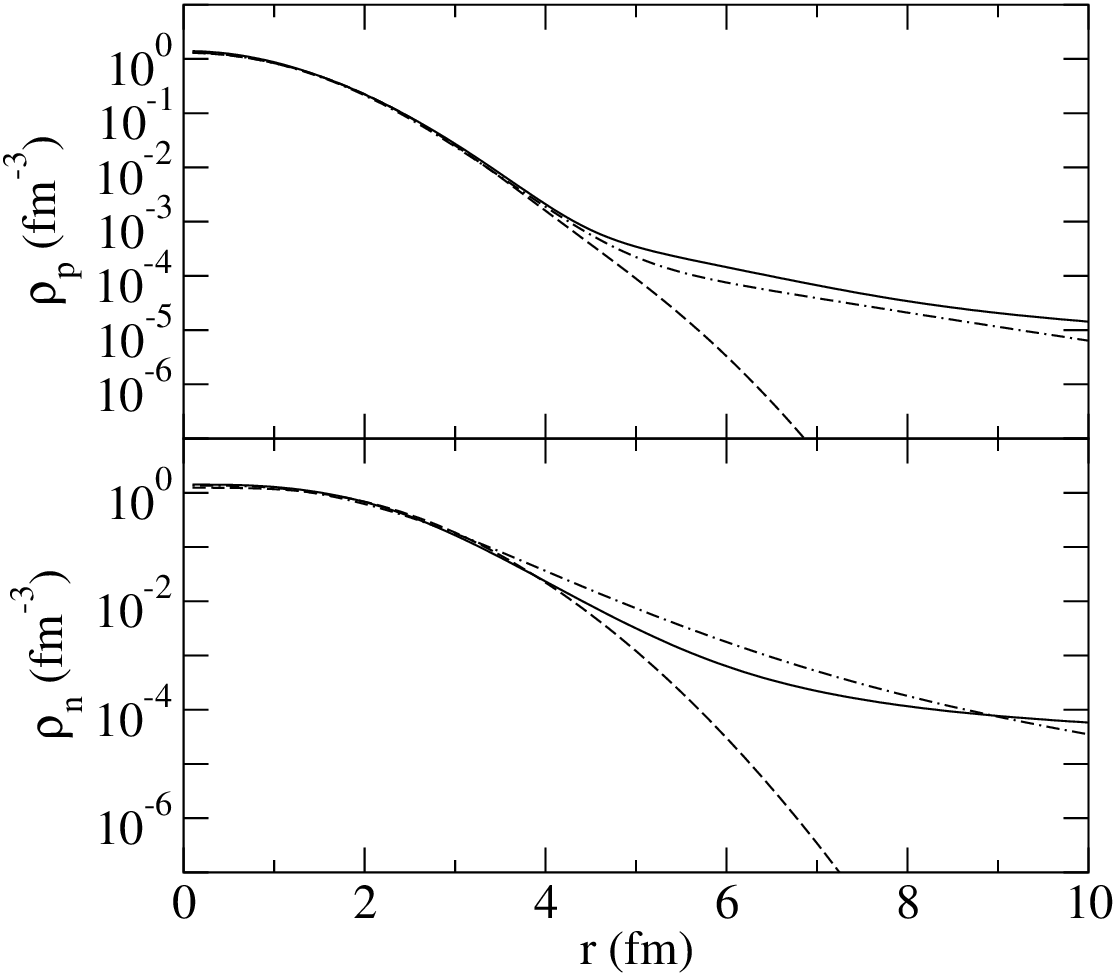}}
\caption{\label{He8-dens}  The  proton   (top)  and  neutron  (bottom)
densities for $^8$He. The HO, LST, and WS results are portrayed by the
dashed, solid, and dot-dashed curves, respectively.}
\end{figure}

We consider $^8$He a test case since it is reasonably well established
that this  nucleus does  not have a  neutron halo. Rather,  the excess
neutron number creates a  skin, whose properties have been established
in analyses  of proton and heavy-ion scattering  data (\cite{Ka00} and
references  cited  therein). Starting  with  the  shell model  results
(OBDME and SP wave functions)  and an oscillator length of 1.6~fm, the
density  profiles for  $^8$He  given the  binding  energies listed  in
Table~\ref{spdata} are shown  in Fig.~\ref{He8-dens}. Proton (neutron)
densities are shown in
the top  (bottom) segment with those  found using the HO,  WS, and LST
functions  displayed  by  the   dashed,  dot-dashed  and  solid  lines
respectively. As with  the WS and LST densities  in $^6$He, extensions
of both  the neutron and  proton densities are observed,  although the
neutron densities are  not as strong at 10~fm as  with $^6$He. This is
consistent with the understanding of  $^6$He having a neutron halo and
$^8$He having a  neutron skin. Note also that the  results for the rms
radius and  reaction cross  section for $^8$He  obtained from  the LST
model are also consistent with a neutron skin description of $^8$He.

\subsection{The case of $^{11}$Li -- a two $s$ orbit problem}

For the  case of $^{11}$Li,  the shell model  calculations \cite{Ka95}
give   dominant   occupancies   for   the  orbitals   as   listed   in
Table~\ref{occups}  and  the  binding  energies  that  were  taken  to
calculate  WS bound  state functions  with  which a  neutron halo  was
created~\cite{La01}  are listed  in Table~\ref{spdata}.  In  this case
there are two $s$ orbitals and the schemes used previously to define a
``halo'' did  not retain orthogonality  of those orbits, nor  does the
LST process set out above. But that can be rectified.

The case  where there are several  orbitals with the same  $\{ ljm \}$
quantum numbers  can be  handled as follows.  Assume, for the  sake of
argument, that there are  three $s_{1/2}$ orbitals, namely $0s_{1/2}$,
$1s_{1/2}$,  and   $2s_{1/2}$,  with  respective   (positive)  binding
energies   $\varepsilon_0  >   \varepsilon_1  >   \varepsilon_2$,  and
corresponding parameters  $\gamma_0 > \gamma_1  > \gamma_2$, according
to  Eq.~(\ref{gamdef}).   Then  LSTs  parameterized   independently  by
$\gamma_0$, $\gamma_1$,  and $\gamma_2$ convert the  HO functions into
orbitals  $\left|  0  \right\rangle$,  $\left| 1  \right\rangle$,  and
$\left| 2 \right\rangle$ which  are normalized but are not orthogonal.
It is a trivial matter  to subtract from $\left| 1 \right\rangle$ that
amount of $\left| 0  \right\rangle$ necessary to regain orthogonality,
and  to  renormalize  the   resultant  new  orbit  vector  $\left|  1'
\right\rangle$.  Notice that  this resultant  state will  have  a long
range aspect  still driven  by $\gamma_1$ since  the subtraction  of a
component proportional  to $\left| 0 \right\rangle$  contains a (much)
shorter range  tail driven  by $\gamma_0.$ In  turn, it is  trivial to
orthogonalize $\left| 2 \right\rangle$ to $\left| 0 \right\rangle$ and
$\left| 1'  \right\rangle$ and renormalize the result  into an orbital
$\left|  2' \right\rangle$;  the tail  of which  is still  governed by
$\gamma_2$. This process is iterative.

\begin{figure}
\scalebox{0.6}{\includegraphics*{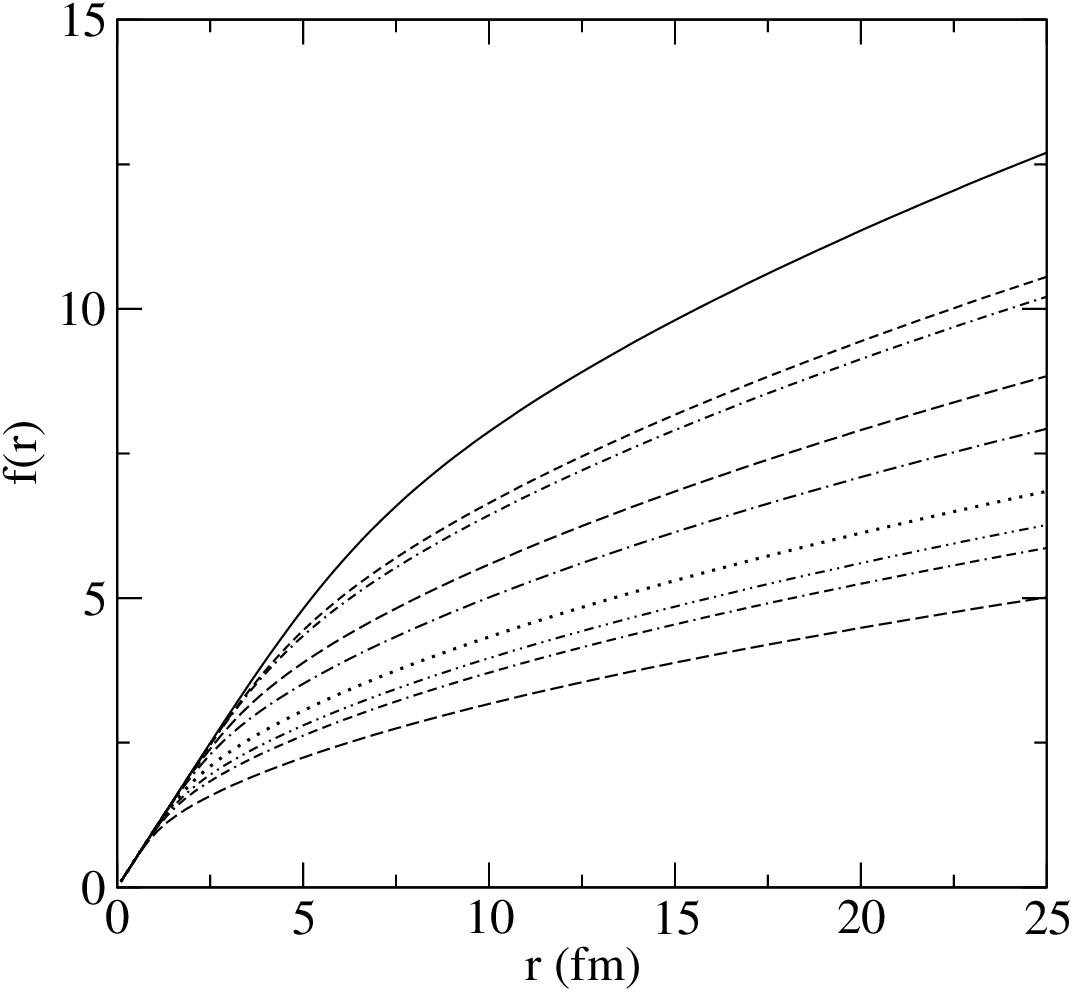}}
\caption{\label{Li11-LST}  The LSTs for  states in  $^{11}$Li obtained
using   the  binding  energies   of  Table~\ref{spdata}.   Curves  are
identified in the text.}
\end{figure}

The LST functions for the set of binding energies for $^{11}$Li listed
in   Table~\ref{spdata},   and   for   $m   =   8$,   are   shown   in
Fig.~\ref{Li11-LST}.  For the  $0s$ case the
proton  and neutron  transform function  is identical  and is  the top
(solid)  curve  in  this  figure.  The  transform  functions  for  the
$0p$-shell are different as indicated by the shallower binding for the
neutrons.    The   functions    for    the   $0p_{\frac{3}{2}}$    and
$0p_{\frac{1}{2}}$ orbits are represented by the dashed and dot-dashed
lines, respectively. The  higher lying set are those  for the protons.
The  remaining curves  are the  transforms  for the  proton: the  $sd$
states  as shown  in  descending sequence  for the  $1s_{\frac{1}{2}}$
(dotted),   the   $0d_{\frac{5}{2}}$   (double-dot-dashed),  and   the
$0d_{\frac{3}{2}}$  (dot-double-dashed)   proton  states.  The  lowest
(long-dashed) curve  is the transform  function for the  three $1s-0d$
neutron states as each was chosen  to have a binding energy of 0.8~MeV
in these calculations.

\begin{figure}
\scalebox{0.6}{\includegraphics*{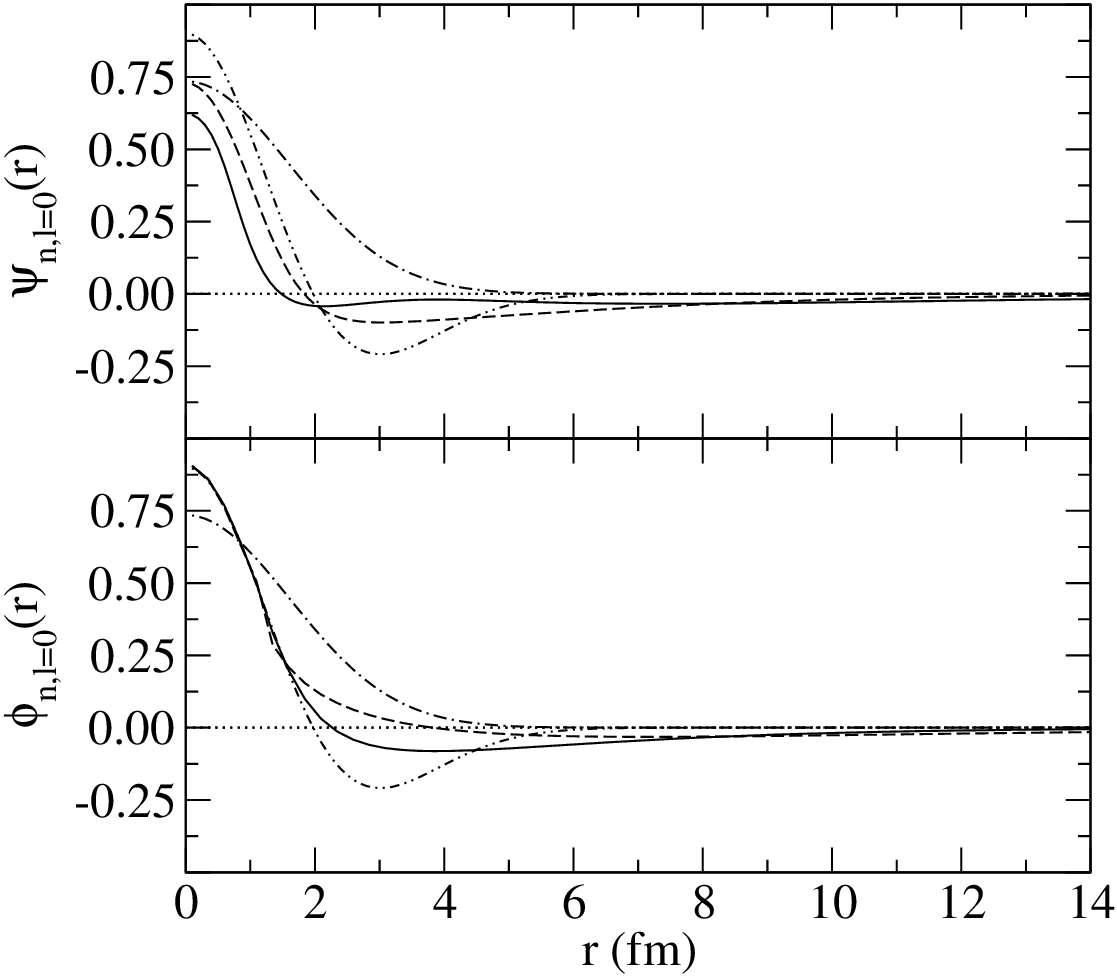}}
\caption{\label{Li11-8hm-0s1sfns}    The   $s$-wave    functions   for
$^{11}$Li. Harmonic  oscillator functions are shown  by the dot-dashed
$(0s)$ and  double-dot-dashed $(1s)$  curves in both  segments. Proton
and neutron LST-transformed $1s$ functions  are given by the solid and
dashed lines,  respectively. The use of  the LST alone  gives the wave
functions   displayed    in   the   bottom    segment   which,   after
orthogonalization, become those given in the upper segment.}
\end{figure}

The  $s$-state wave functions  that result  after re-orthogonalization
are  shown  in Fig.~\ref{Li11-8hm-0s1sfns}.   Because  the proton  and
neutron $0s$  orbits were both chosen  to be bound by  33~MeV there is
little change  to their  wave functions from  that of the  starting HO
function. Of course the long  range form differs: the transformed wave
functions have an exponential character  while the HO is Gaussian. But
those  differences   cannot  be  discerned  on  the   scales  used  in
Fig~\ref{Li11-8hm-0s1sfns}. Those $0s$ HO  wave functions are shown in
both the upper and lower parts of this figure by the dot-dashed curve.
The $1s$ states change markedly not only by virtue of the LST but also
with re-orthogonalization. Due to  the LST alone, wave functions shown
in  the  bottom  panel  of Fig.~\ref{Li11-8hm-0s1sfns}  result.  After
orthogonalization,  the  wave functions  displayed  in  the top  panel
result. In both panels  the transformed $1s$ wave functions determined
with a  binding energy  of 2.8~MeV (proton)  and by  0.8~MeV (neutron)
binding energy are shown by  the solid and dashed curves respectively.
Not only are the spatial  variations of the transformed wave functions
quite  different from  that  of the  initial  $1s$ oscillator  (bottom
panel)  as the  transform varies  the HO  to get  the  relatively weak
binding form of  the exponentials, but also those  changes are altered
with the central  radial values of the LST  functions markedly reduced
under the  constraint that the  $0s$ and $1s$ results  be orthonormal.
Indeed both the proton and  neutron $1s$ orbit functions are extended,
though by virtue of its weaker binding the neutron one is the more so.
Then with the rather large occupancy of neutrons in the $1s$ orbit the
neutron matter profile has the character of a neutron halo.

\begin{figure}
\scalebox{0.6}{\includegraphics*{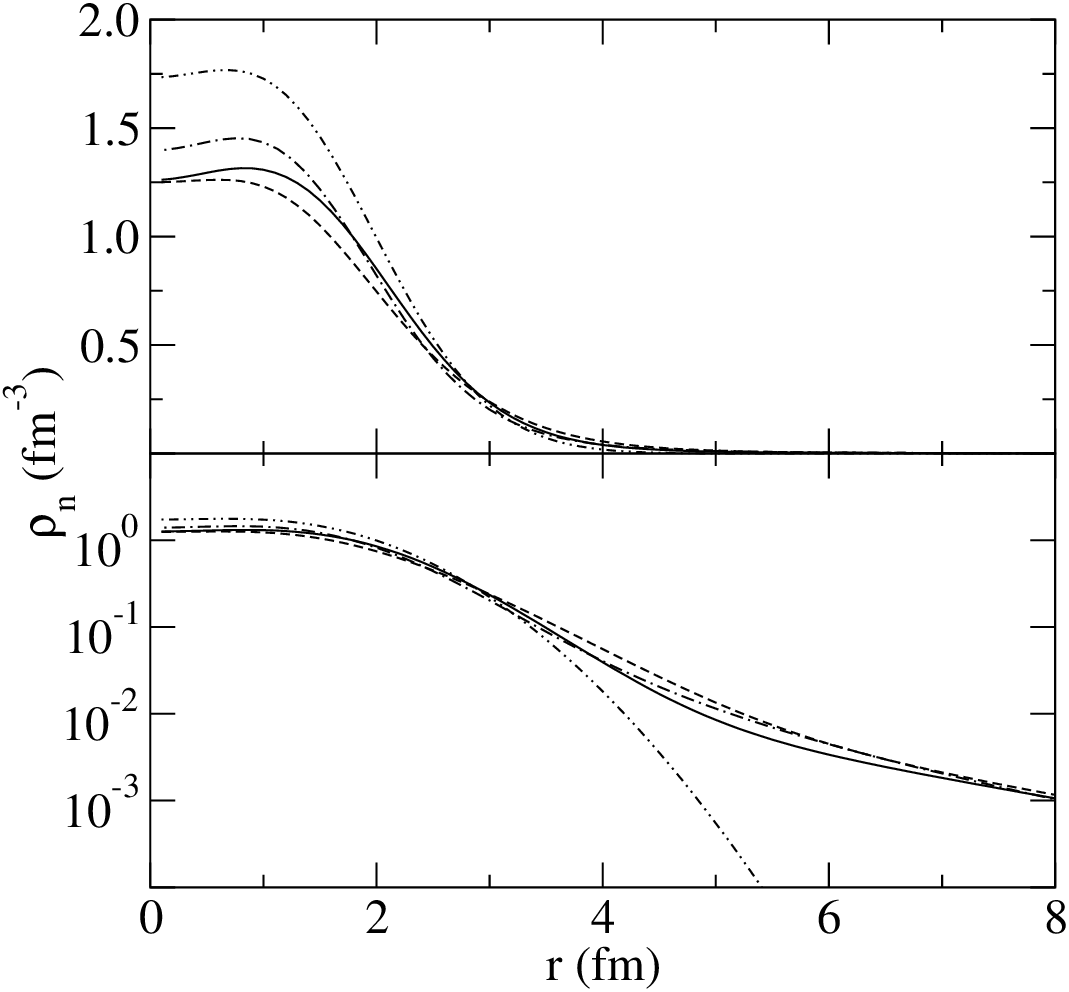}}
\caption{\label{Li11_neutron}   The   neutron   matter  densities   in
$^{11}$Li given  by the HO  (double-dot-dashed), the $m =  8$ harmonic
mean  transformations (solid),  the WS  (dot-dashed) and  the $m  = 4$
harmonic mean transformations (dashed) models.}
\end{figure}

Diverse neutron matter densities are shown in Fig.~\ref{Li11_neutron}.
They  are  the HO  result  (double-dot-dashed  curve),  the WS  result
(dot-dashed curve), and the  two harmonic mean transform results; that
for $m = 8$ (solid curve) and  for $m = 4$ (dashed curve). The neutron
densities are shown  in a linear plot (top)  and in a semi-logarithmic
plot  (bottom)   to  stress  the  short  and   long  range  properties
differently.  Clearly, the  power used  in the  harmonic mean  form of
transform makes a  significant difference. The $m =  4$ transforms all
vary from  the linear limit condition  at a rather  small radius since
the resulting  wave functions  are reduced to  effect the  quite small
value of the central neutron density. As with the He isotopes, the LST
densities are  more similar to those  obtained from the  WS model. The
main difference lies  near the centre; the WS  density is higher. That
is compensated  by a sharper fall-off  compared to the LST  up to 4~fm
after which  both the  LST and WS  results exhibit a  somewhat similar
extension compared to the HO density.

\subsection{A stable nucleus -- $^{40}$Ca}

The LST  scheme should be  appropriate also in dealing  with structure
assumed  for stable  nuclei. Notably  its use  should not  vitiate any
success that  has been  achieved to date  when basic  model structures
have   been   used   as   input   in   studies   of   proton   elastic
scattering~\cite{Am00}. As  a test we consider the  case of $^{40}$Ca,
the structure  of which has been  determined by both  a standard shell
model   approach~\cite{Ka01}  and   by  a   Skyrme-Hartree-Fock  (SHF)
prescription~\cite{Br98}.   Within   the   oscillator   model,   while
Karataglidis  and Chadwick~\cite{Ka01}  used an  oscillator  length of
2.0~fm, we found that better scattering results were obtained with the
shell model  wave functions by allowing  a small reduction  of that to
1.9~fm.

\begin{figure}
\scalebox{0.6}{\includegraphics*{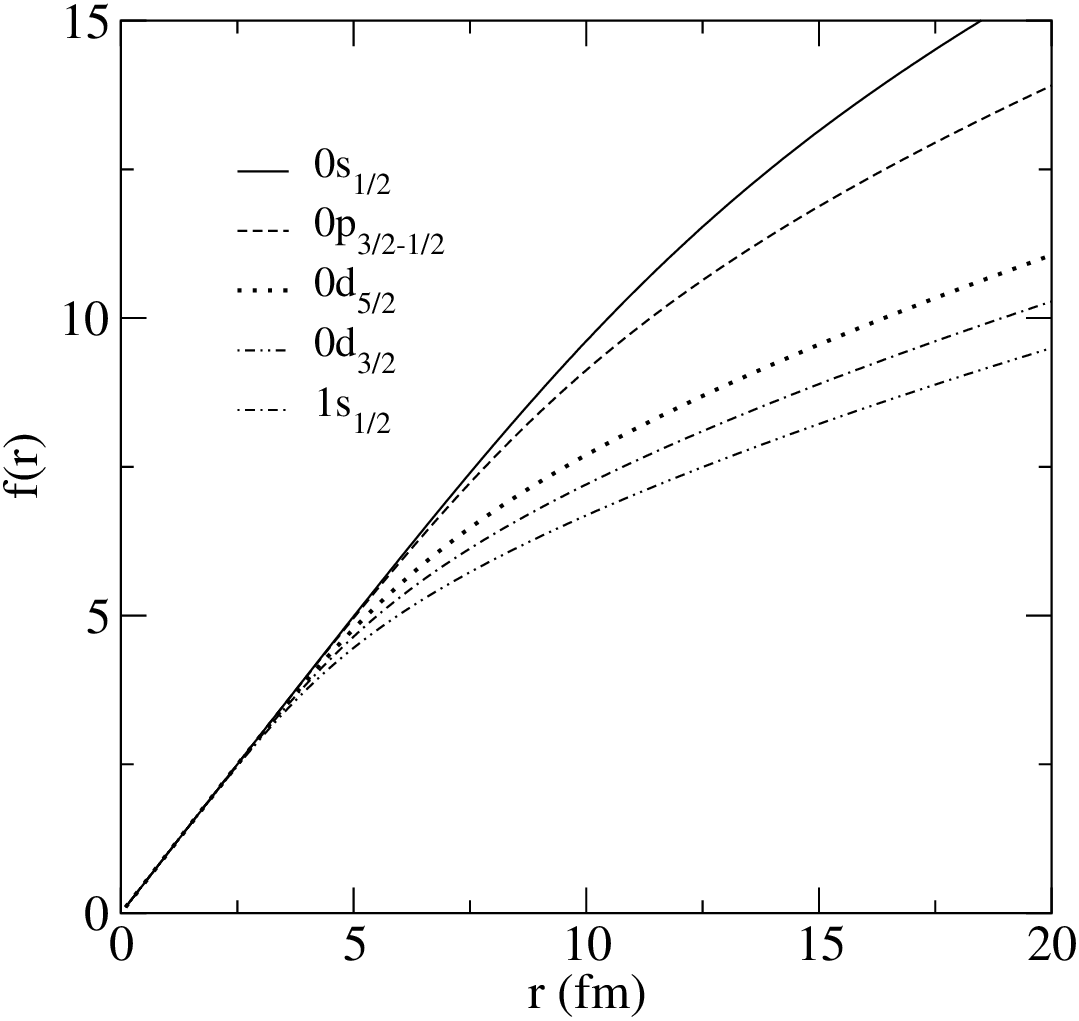}}
\caption{\label{Ca40-f} The $m = 8$ harmonic mean coordinate transform
functions for the orbits of $^{40}$Ca found using the binding energies
listed in  Table~\ref{Ca40-bes}. The transforms for the  six orbits of
the $0s$, $0p$, and $1s-0d$ shells are as identified in the legend.}
\end{figure}

\begin{figure}
\scalebox{0.6}{\includegraphics*{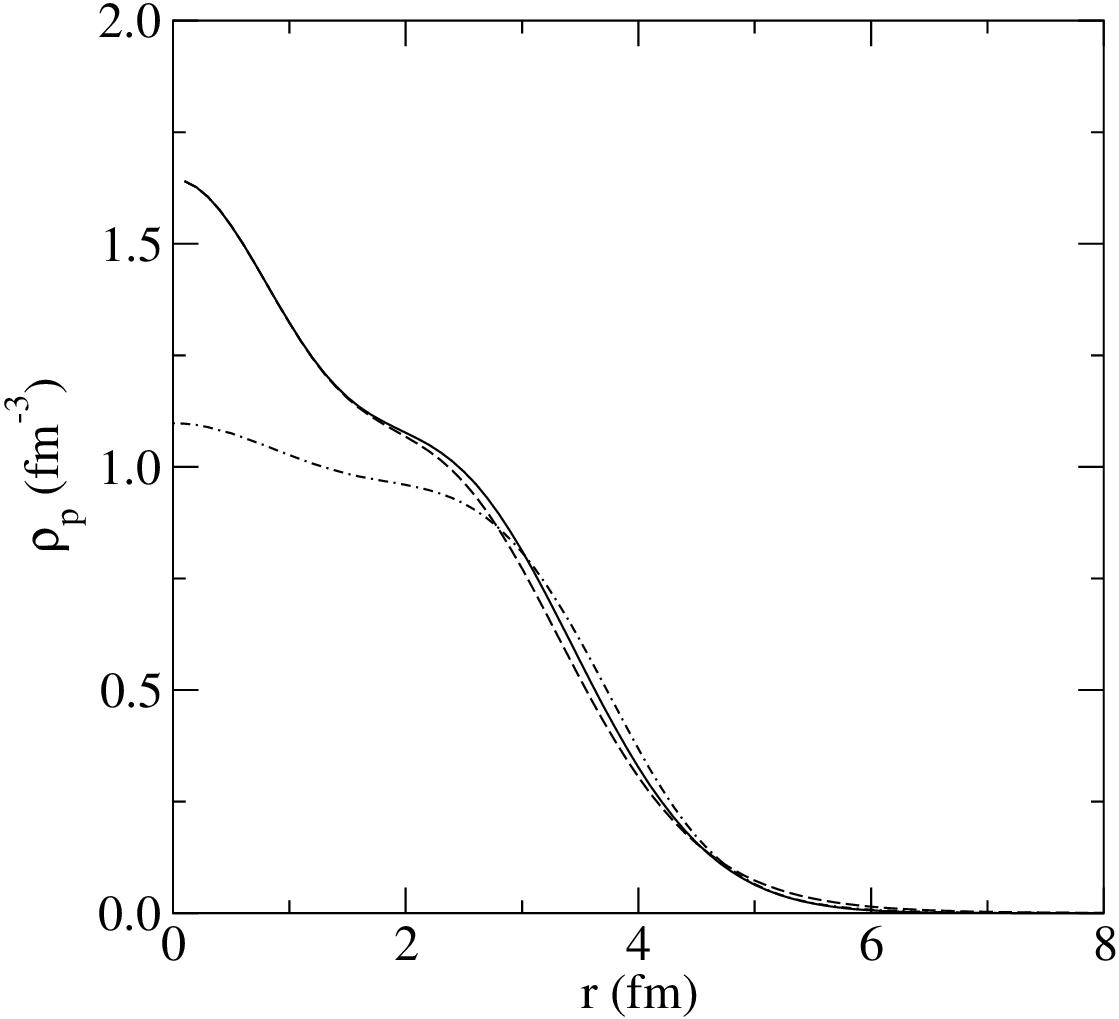}}
\caption{\label{Ca40-dens-HO+LST}  The   proton  matter  densities  in
$^{40}$Ca  given  by  the HO  (dashed),  the  $m  = 8$  harmonic  mean
transformations (solid), and the SHF/SKX (dot-dashed) models.}
\end{figure}

We applied the  LST to the shell model wave functions  to obtain a new
set with exponential tails consistent with the binding energies listed
in Table~\ref{Ca40-bes}.   In that table  we also give the  rms values
for each occupied orbit.
\begin{table}
\begin{ruledtabular}
\caption{\label{Ca40-bes}
Adopted binding energies (MeV) for nucleon orbits
in $^{40}$Ca and their rms radii.}
\begin{tabular}{cccccc}
 Orbit & B.E. (proton) & B.E. (neutron) & rms$_{\rm HO}$
($b$ = 1.9 fm) & rms (proton) & rms (neutron) \\
\hline 
$0s_{\frac{1}{2}}$ & 67.0 & 67.0 & 2.33 & 2.33 & 2.33 \\
$0p_{\frac{3}{2}}$ & 39.2 & 39.2 & 3.00 & 3.01 & 3.01 \\
$0p_{\frac{1}{2}}$ & 39.0 & 39.0 & 3.00 & 3.01 & 3.01 \\
$0d_{\frac{5}{2}}$ & 21.7 & 15.3 & 3.55 & 3.61 & 3.59 \\
$0d_{\frac{3}{2}}$ & 15.3 & \ 8.3 & 3.55 & 3.66 & 3.99 \\
$1s_{\frac{1}{2}}$ & 17.9 & 11.4 & 3.55 & 3.66 & 3.81 \\
\end{tabular}
\end{ruledtabular}
\end{table}
Clearly, with regard to the rms radii of each orbit, the modulation of
the long range character of the Gaussians is not severe as the binding
energies are all reasonably large.   In all cases the transform radius
($r_t$) is quite large as is evident in Fig.~\ref{Ca40-f}.
From this figure, the linearity of  all of the transform terms is well
retained to  near 4~fm, by which  distance the matter  density is less
than  10\% of  its central  value.  Thus  we do  not expect  any major
difference in results obtained using the LST functions in calculations
from  those  found  when  the  Gaussians themselves  are  used.   That
expectation  is   heightened  by   a   study  of  the matter  density.
Considering   the   proton   distributions   only,   we   compare   in
Fig.~\ref{Ca40-dens-HO+LST}, the results obtained from the shell model
($b = 1.9$~fm), from the  LST functions deduced from that shell model,
and by  that given by the  SHF/SKX description of the  ground state of
$^{40}$Ca.
In this  case, the LST density is  similar to that of  the input shell
model function; the surface being slightly extended.  Both differ most
noticeably from the  SHF/SKX in the nuclear interior,  and the SHF/SKX
model  density   extends  further  still.   But   the  large  interior
difference is not very important in the analyses we make as the volume
integral contribution  of that region is  not large.  In  use of these
wave functions  to define  optical potentials that  volume integration
contribution  as well  as  the inherent  absorption  makes the  region
inside  about 2~fm  of small  import for  most scattering.  It  is the
surface differences that one may expect to most influence results.

\section{Applications in scattering analyses}
\label{results}

The  harmonic mean  LST wave  functions determined  from the  $m  = 8$
formulation  have been  used  as input  into  calculations of  elastic
scattering of the radioactive ions from hydrogen targets as well as of
proton  scattering  from  the  stable nucleus  $^{40}$Ca.  A  modified
version of the code  DWBA98~\cite{Ra98} has been used with appropriate
effective  $NN$ interactions  for  each energy  considered with  OBDME
obtained for each nucleus as outlined earlier.

\subsection{Scattering of $^{6,8}$He and $^{11}$Li}

Elastic  scattering of  $24.5A$, $40.9A$  and $70.5A$~MeV  $^6$He ions
from     hydrogen      has     been     measured      and     analyzed
\cite{La01,St02,Ko93,Ko97} revealing  that this nucleus  has a neutron
matter distribution more consistent with a neutron halo than a neutron
skin, as the naive shell  model predicts. That was definitely the case
considering the  $24.5A$~MeV elastic scattering  data. At $40.9A$~MeV,
the DWA analysis of the  scattering data for excitation of the $2^+_1$
state  was the  prime evidence  for  a halo.  The $70.5A$~MeV  elastic
scattering data  do not  extend to large  enough momentum  transfer to
distinguish the halo aspect clearly but we include it to show a set of
data  for which  the  method used  to  predict the  cross sections  is
reliable. In those previous  studies the neutron halo was artificially
created by choosing weak binding for the $0p$ neutron orbits and using
WS potentials to define the radial wave functions.

\begin{figure}
\scalebox{0.6}{\includegraphics*{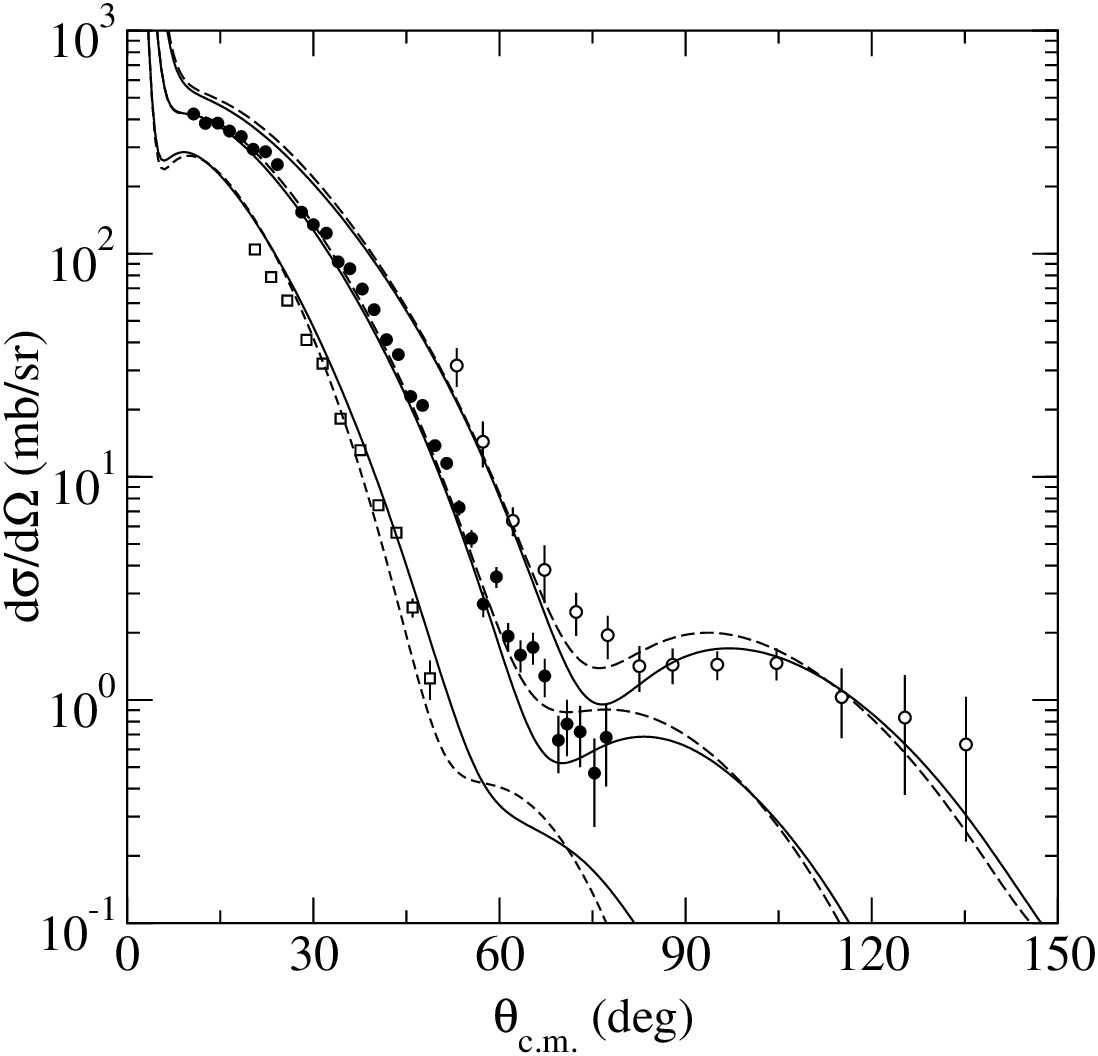}}
\caption{\label{He6-8hm-cs} The  elastic scattering differential cross
sections  for  $24.5A$,  $40.9A$,  and $70.5A$~MeV  $^6$He  ions  from
hydrogen. For each  energy the LST ($m=8$) result  is portrayed by the
solid line while that result obtained from the WS SP wave functions is
given by the dashed line.}
\end{figure}

Differential cross  sections for  the scattering of  $24.5A$, $40.9A$,
and $70.5A$~MeV $^6$He ions from  Hydrogen, as obtained using the $m =
8$  LST  on   the  HO  functions  with  binding   energies  listed  in
Table~\ref{spdata},    are   shown    by   the    solid    curves   in
Fig.~\ref{He6-8hm-cs}. Therein the data are shown by the open circles
($24.5A$~MeV), the filled circles  ($40.9A$~MeV), and the open squares
($70.5A$~MeV).  The previous  halo  results \cite{La01,St02,Ka00}  are
shown  by  the dashed  curves  for  comparison.  Our transformed  wave
functions serve to  correct the description of these  data as does the
more \textit{ad  hoc} selection of disparate  WS functions \cite{La01}
for the  occupied orbits by  being distinctively different  from those
obtained when no extension to  neutron matter was considered. The data
at $24.5A$  and $40.9A$~MeV extend  beyond the first minimum  into the
region where one may distinguish between the halo and non-halo regions
\cite{La01,St02}. Both  the WS and  LST results agree well  with those
data, indicating that our modifications  to the HO wave functions with
the LST  make the necessary corrections  to explain the  data. This is
consistent with  our results  concerning both the  rms radius  and the
reaction cross section. Both the WS and LST results agree equally well
with the available data at $70.5A$~MeV.

\begin{figure}
\scalebox{0.6}{\includegraphics*{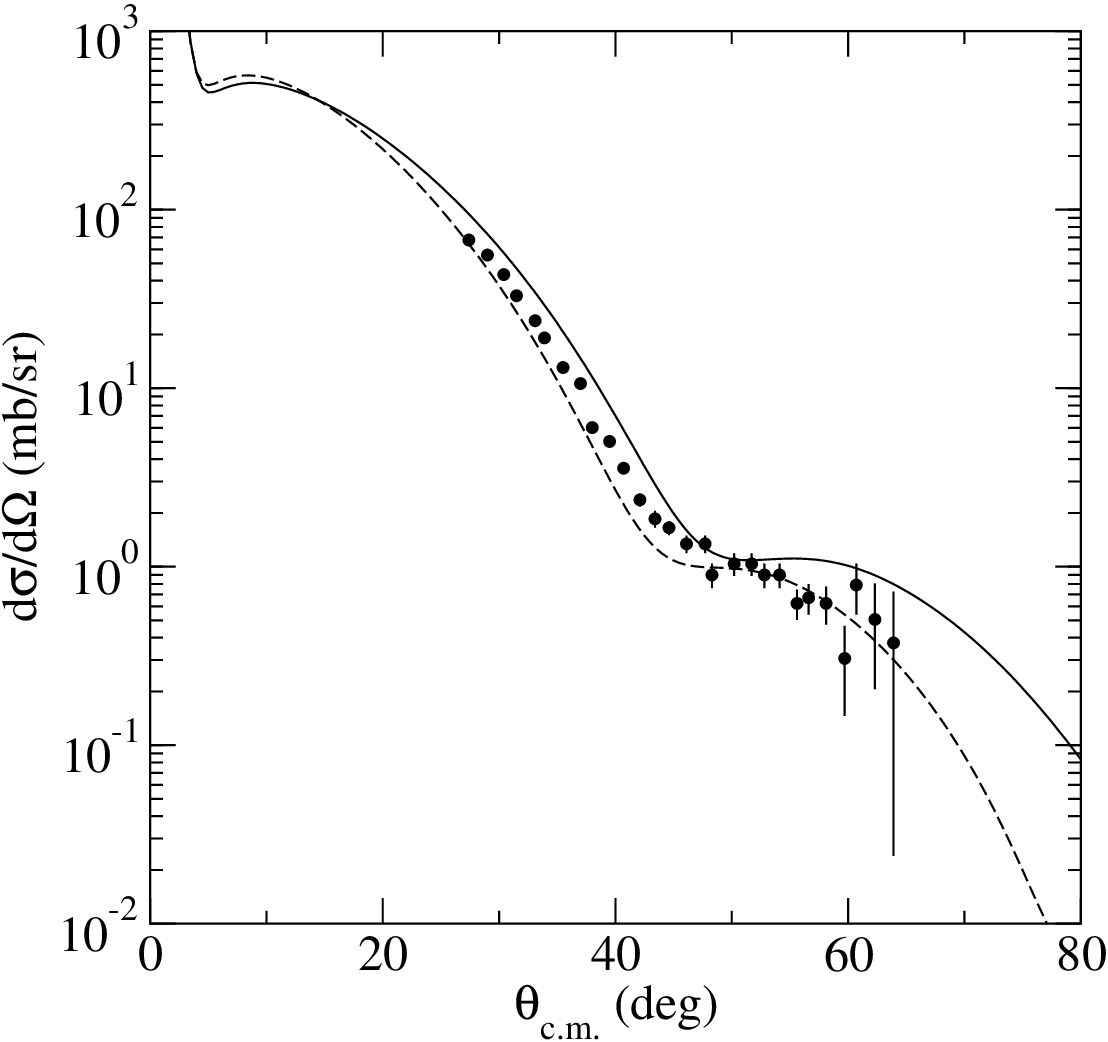}}
\caption{\label{He8-cs}  The  elastic  scattering  differential  cross
sections for $72A$~MeV  $^8$He ions from hydrogen. The  results of the
calculations  made  using the  LST  ($m=8$) and  WS  sets  of SP  wave
functions are displayed by the solid and dashed lines, respectively.}
\end{figure}

Previously  the  scattering  data  analyses confirmed  what  had  been
expected from  heavy ion collision  studies that $^8$He has  a neutron
skin but not  the extended distribution one now  identifies as a halo.
The  appropriate LST  for SP  wave functions  for this  nucleus, again
predicated  upon  an  oscillator  length  of 1.6~fm,  retains  a  skin
attribute and results in  the differential cross section for $72A$~MeV
$^8$He  ions from  hydrogen shown  in Fig.~\ref{He8-cs}  by  the solid
line. The data  were taken from Refs. \cite{Ko93,Ko97}  and the dashed
curve  is the  result that  was obtained  previously  \cite{Ka00} when
those SP wave functions were taken as the earlier published WS set.
Both results do well in describing the available data.

\begin{figure}
\scalebox{0.6}{\includegraphics*{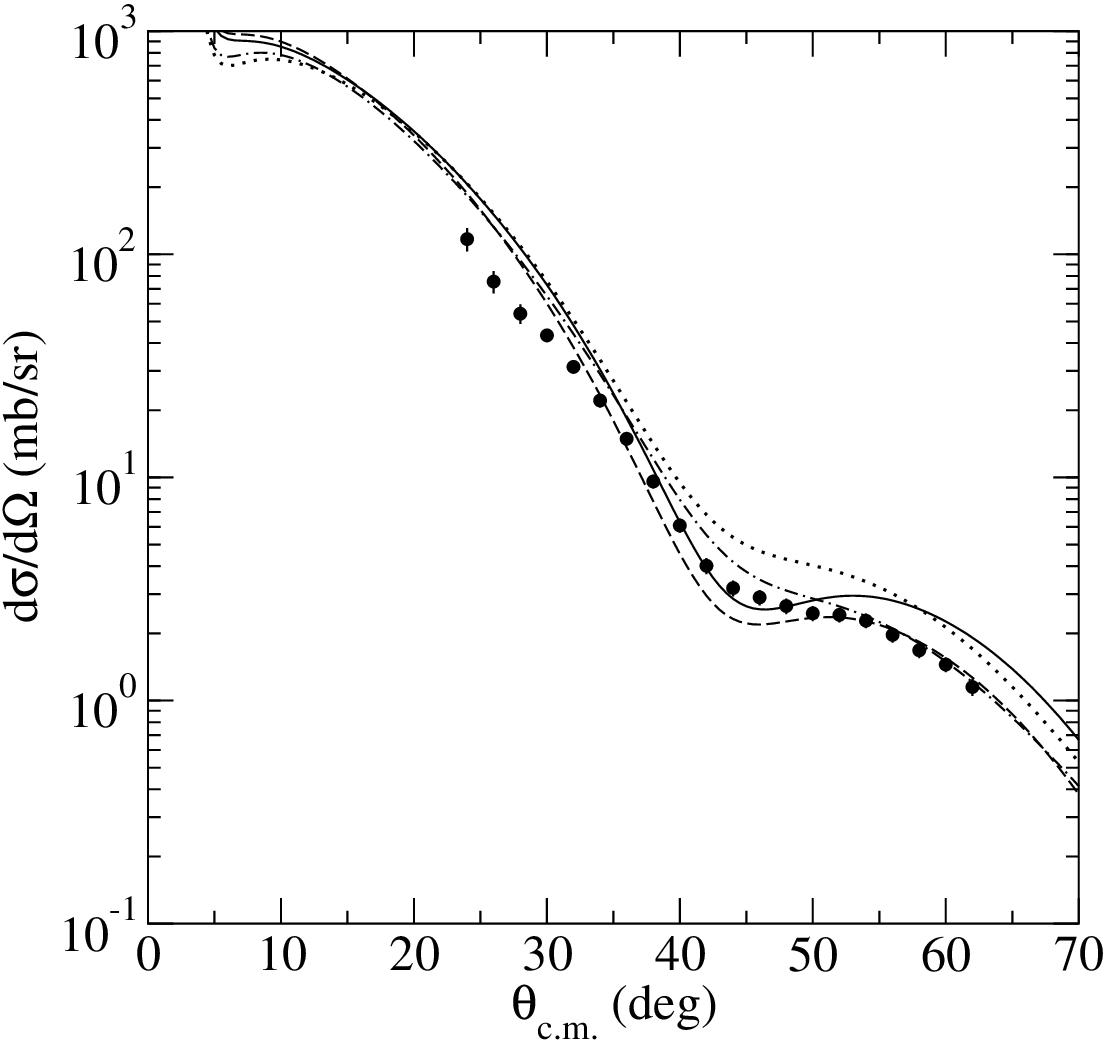}}
\caption{\label{li11-xsec}  Data   from  the  elastic   scattering  of
$62A$~MeV  $^{11}$Li ions  from hydrogen  compared to  the predictions
made using the  basic shell model wave functions  (dotted curve), with
the  WS (halo)  functions (dot-dashed  curve), and  with the  $m  = 4$
(dashed curve) and $m=8$  (solid curve) harmonic mean transformed wave
functions.}
\end{figure}

The  nucleus $^{11}$Li  is known  to have  an extended  neutron (halo)
density. That was confirmed from the analyses of elastic scattering of
$^{11}$Li  ions  from  hydrogen  \cite{Ka00}  and  those  results  are
displayed  again in  Fig.~\ref{li11-xsec} along  with our  new results
obtained by  using the same  approach, with the same  effective force,
but with the transformed HO wave functions.
As  noted previously the  difference between  using WS  wave functions
with binding  energies chosen to obtain  a halo in  the neutron matter
density in this  nucleus and those wave functions that  set it to have
only a skin, is striking. Only with the halo specification does a good
prediction of the  data result. That is true  also for the transformed
HO functions,  with the $m=4$ LST providing  excellent reproduction of
the data. While the results  obtained from the LST transformation with
$m=8$ provides  good reproduction up to  $50^{\circ}$ it overestimates
the larger angle data.

\subsection{Scattering from  $^{40}$Ca}

\begin{figure}
\scalebox{0.6}{\includegraphics*{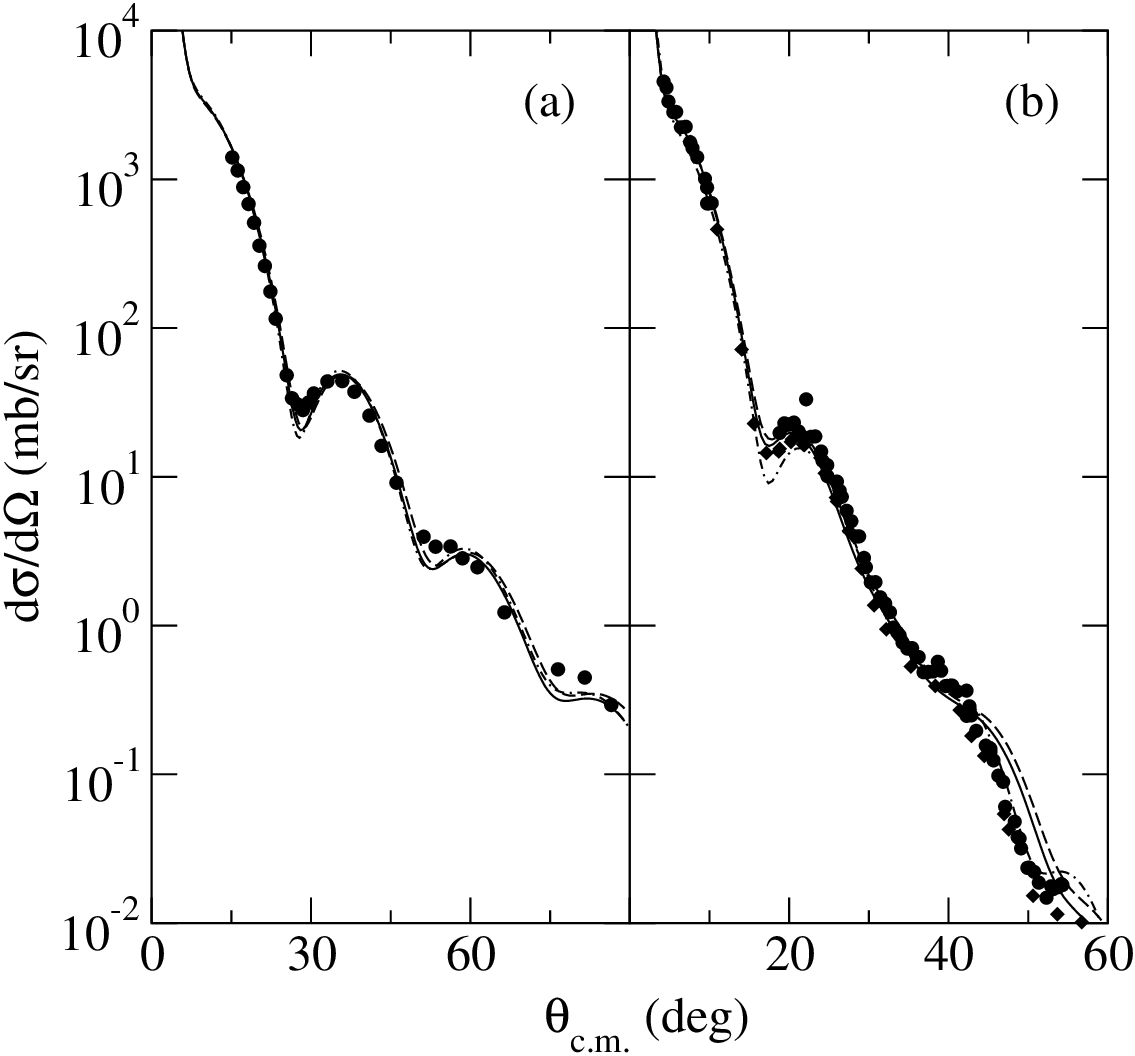}}
\caption{\label{Ca40-LST-cs}  Data  from  the  elastic  scattering  of
65~MeV (a)  and of  200~MeV (b) protons  from $^{40}$Ca  compared with
predictions made  using the basic  shell model wave  functions (dashed
curve), with the SHF/SKX functions (dot-dashed curve), and with the $m
= 8$ (solid curve) harmonic mean transformed wave functions.}
\end{figure}

Finally  we consider the  use of  the LST  functions for  $^{40}$Ca in
generating  optical  potentials. With  those  functions  we have  made
predictions of  the elastic scattering  of 65 and of  200~MeV protons.
Such   data  were   analyzed  recently   \cite{Ka02}  and   very  good
differential  cross-section results were  obtained for  both energies;
especially  when the  SHF/SKX model  wave functions  were  used. Those
SHF/SKX  results are  shown again  in Fig.~\ref{Ca40-LST-cs}  for both
energies by the dot-dashed curves.
The  shell model ($b  = 1.9$~fm)  results are  those portrayed  by the
dashed curves  while the LST function  results are given  by the solid
curves.  Note that the shell model results are varied from those found
earlier~\cite{Ka02}; the result of  our changing the oscillator length
slightly from  that defined by  Karataglidis and Chadwick~\cite{Ka02}.
The  adjustment was  made  specifically to  obtain  the best  possible
agreement with the data from the shell model. That allows for the most
sensitivity to changes wrought by the LST.  The changes are slight but
they  in fact  improve agreement  with observation.   But  neither our
shell  model or the  LST built  from it  give results  as good  as the
SHF/SKX  model of  structure.  Clearly  while  the LST  may give  more
reasonable matter profiles  to a model of the  ground state structure,
it is not a  panacea for a too limited initial guess.   Use of the LST
approach with ``best model'' structures of nuclei are in train.


\section{Summary and conclusions}

Using local scale transformations  of the radial coordinate within the
Gaussian  wave  functions  assumed  to  describe  bound  nucleons   in
shell model studies, gives new descriptions of those nucleon functions
that  have  exponentially decreasing  forms  consistent with  selected
values   for  their   binding  energies.    Orthonormality   of  those
transformed  wave functions can  be assured  quite easily.   Herein we
have considered an harmonic mean form of local scale transforms.

As an empirical example, the harmonic mean LST (of rank 8) was used to
specify  a set  of  single nucleon  bound  state orbitals  for use  in
defining  optical potentials  to  describe the  elastic scattering  of
light  mass radioactive ions  ($^{6,8}$He and  $^{11}$Li specifically)
from hydrogen as well as for the scattering of protons from the stable
nucleus   $^{40}$Ca.   Those  optical   potentials   were  formed   by
$g$-folding: folding  of complex effective $NN$  interactions in which
medium modification due  to both Pauli blocking and  a background mean
field had been taken into  consideration with the LST generated single
nucleon  wave functions  weighted by  the OBDME  given by  shell model
calculations.  The resultant nonlocality  of those  optical potentials
was  treated  exactly.  The  results  for the  elastic  scattering  of
$24.5A$, $40.9A$  and of $70.5$~MeV  $^6$He ions from  hydrogen agreed
well with  both the  data and previous  calculations in which  WS wave
functions  were used. With  both the  \textit{ad hoc}  WS and  the LST
formed wave  functions, $^6$He has a neutron  distribution so extended
from that associated  with the shell model (Gaussian  functions) as to
be consistent  with a halo. Notably,  the WS and LST  densities are in
good agreement.  However, it is also  of note that in  order to obtain
such  agreement  in the  densities,  the  chosen  sets of  SP  binding
energies need  not be the same,  as the underlying  potentials (WS and
HO) are  different. For  $^8$He, the LST  (and WS)  functions involved
also give  good results  in comparison with  scattering data  taken at
$72A$~MeV. In this case the neutron  extension is not as large as that
for $^6$He  resulting in this nucleus  defined to have  a neutron skin
rather than a  halo. Yet with both nuclei we find  an extension of the
proton  density beyond  the HO  result.  That dilution  of the  proton
density  is influenced  by the  extension of  the neutron  density, as
expected for neutron-rich nuclei. Also, a good result in comparison to
data is  obtained when  cross sections for  the elastic  scattering of
$62A$~MeV $^{11}$Li  ions from hydrogen were considered.  The LST wave
functions again  extend the neutron  distribution for this  nucleus so
much that  we deem it to  be a neutron halo.  In this case  it was due
mainly  to the  neutron  occupancy of  the  $0p_{1/2}$ and  $1s_{1/2}$
orbits and we took care to ensure that $1s_{1/2}$ orbit was orthogonal
to the $0s_{1/2}$. In this case we noted that the rank of the harmonic
mean has  some import  regarding the quality  of the agreement  of the
results with data.

Finally, as  a check  case, we found  that using  the LST to  vary the
shell model  single nucleon wave  functions for a stable  nucleus case
did not vitiate the good results previously found for scattering cross
sections with  potentials formed by  $g$-folding with the  shell model
wave  functions themselves.  The cross  sections from  65  and 200~MeV
protons   elastically  scattered   from  $^{40}$Ca   were  considered.
$^{40}$Ca was  considered also  as there exist  SHF wave  functions to
describe its ground state and whose use in $g$-folding gave potentials
and  scattering results  also in  very good  agreement with  the data.
However,  the densities formed  by the  shell and  the SHF  models are
different,  most noticeably  in the  central  region and  also in  the
surface.  The LST  modifications  vary the  shell  model density  most
through the surface region and  therefore does not give changes inside
the  nucleus to  match the  SHF results.  Of course,  as the  SHF wave
functions need  not have appropriate  exponential tails either,  it is
feasible  to   apply  the   LST  scheme  to   those.  That   is  under
investigation.

\begin{acknowledgments}
This research  was supported by  a research grant from  the Australian
Research Council.
\end{acknowledgments}

\bibliography{LST_V3}

\end{document}